\documentclass[pra,aps,twocolumn,superscriptaddress,nofootinbib,nopacs]{revtex4} 

\usepackage{amsmath}  \usepackage{amssymb}  \usepackage{amsfonts}  \usepackage{bm}  \usepackage{bbm}  \usepackage{braket}  \usepackage{color}  \usepackage{comment}  \usepackage{dcolumn}  \usepackage{epsfig}
\usepackage{gensymb}  \usepackage{graphicx}  \usepackage{indentfirst}  \usepackage{lmodern}  \usepackage{mathrsfs}  \usepackage{mathtools}  \usepackage{psfrag}  \usepackage{pst-all}  \usepackage{soul}  \usepackage{units}  \usepackage{xcolor}
\usepackage{float} 
\usepackage[colorlinks,linkcolor=blue,citecolor=blue,urlcolor=blue,hyperindex,driverfallback=dvipdfm]{hyperref}  \usepackage[T1]{fontenc} 

\def\ii{{\rm i}}  \def\ee{{\rm e}}
\def\me{m_{\rm e}}  \def\kB{{k_{\rm B}}}
\def\Ree{{\rm Re}}  \def\Imm{{\rm Im}}
\def\Ab{{\bf A}}      \def\bb{{\bf b}}  \def\Eb{{\bf E}}    \def\Fb{{\bf F}}  \def\fb{{\bf f}}            \def\jb{{\bf j}}    \def\kb{{\bf k}}        \def\pb{{\bf p}}    \def\qb{{\bf q}}  \def\Rb{{\bf R}}  \def\rb{{\bf r}}      \def\vb{{\bf v}} 
    \def\zz{\hat{\bf z}}    \def\rr{\hat{\bf r}}      \def\RR{\hat{\bf R}}  
      
\begin{document} 

\def\bibsection{\section*{\refname}} 

\title{Optical Excitations with Electron Beams: Challenges and Opportunities
}


\author{F.~Javier~Garc\'{\i}a~de~Abajo}
\email{javier.garciadeabajo@nanophotonics.es}
\affiliation{ICFO-Institut de Ciencies Fotoniques, The Barcelona Institute of Science and Technology, 08860 Castelldefels (Barcelona), Spain}
\affiliation{ICREA-Instituci\'o Catalana de Recerca i Estudis Avan\c{c}ats, Passeig Llu\'{\i}s Companys 23, 08010 Barcelona, Spain}
\author{Valerio~Di~Giulio}
\affiliation{ICFO-Institut de Ciencies Fotoniques, The Barcelona Institute of Science and Technology, 08860 Castelldefels (Barcelona), Spain}




\begin{abstract}
Free electron beams such as those employed in electron microscopes have evolved into powerful tools to investigate photonic nanostructures with an unrivaled combination of spatial and spectral precision through the analysis of electron energy losses and cathodoluminescence light emission. In combination with ultrafast optics, the emerging field of ultrafast electron microscopy utilizes synchronized femtosecond electron and light pulses that are aimed at the sampled structures, holding the promise to bring simultaneous sub-{\AA}--sub-fs--sub-meV space-time-energy resolution to the study of material and optical-field dynamics. In addition, these advances enable the manipulation of the wave function of individual free electrons in unprecedented ways, opening sound prospects to probe and control quantum excitations at the nanoscale. Here, we provide an overview of photonics research based on free electrons, supplemented by original theoretical insights, and discussion of several stimulating challenges and opportunities. In particular, we show that the excitation probability by a single electron is independent of its wave function, apart from a classical average over the transverse beam density profile, whereas the probability for two or more modulated electrons depends on their relative spatial arrangement, thus reflecting the quantum nature of their interactions. We derive first-principles analytical expressions that embody these results and have general validity for arbitrarily shaped electrons and any type of electron-sample interaction. We conclude with some perspectives on various exciting directions that include disruptive approaches to non-invasive spectroscopy and microscopy, the possibility of sampling the nonlinear optical response at the nanoscale, the manipulation of the density matrices associated with free electrons and optical sample modes, and appealing applications in optical modulation of electron beams, all of which could potentially revolutionize the use of free electrons in photonics.
\end{abstract}

\maketitle 
\tableofcontents 


\section{Introduction}

The last two decades have witnessed spectacular progress in our ability to control light down to deep-subwavelength scales thanks to advances in nanofabrication using bottom-up approaches (colloid chemistry \cite{BCN05} and surface science \cite{CRJ10}) and  top-down techniques (electron-beam \cite{DFB12} (e-beam) and focused-ion-beam \cite{NLO09} lithographies), as well as combinations of these two types of methods \cite{paper335,DJD20}. In parallel, substantial improvements in optics have enabled the acquisition of spectrally resolved images through scanning near-field optical microscopy \cite{HTK02,BKO10,WFM14} (SNOM) and super-resolution far-field optics \cite{BPS06,YZ19}, in which the diffraction limit is circumvented either by relying on nanoscale scatterers ({\it e.g.}, metallic tips \cite{HTK02,BKO10,WFM14}) or by targeting special kinds of samples ({\it e.g.}, periodic gratings \cite{YZ19} or fluorophore-hosting cells \cite{BPS06}). However, light-based imaging is far from reaching the atomic level of spatial resolution that is required to investigate the photonic properties of vanguard material structures.

\begin{figure*}
\centering{\includegraphics[width=1.0\textwidth]{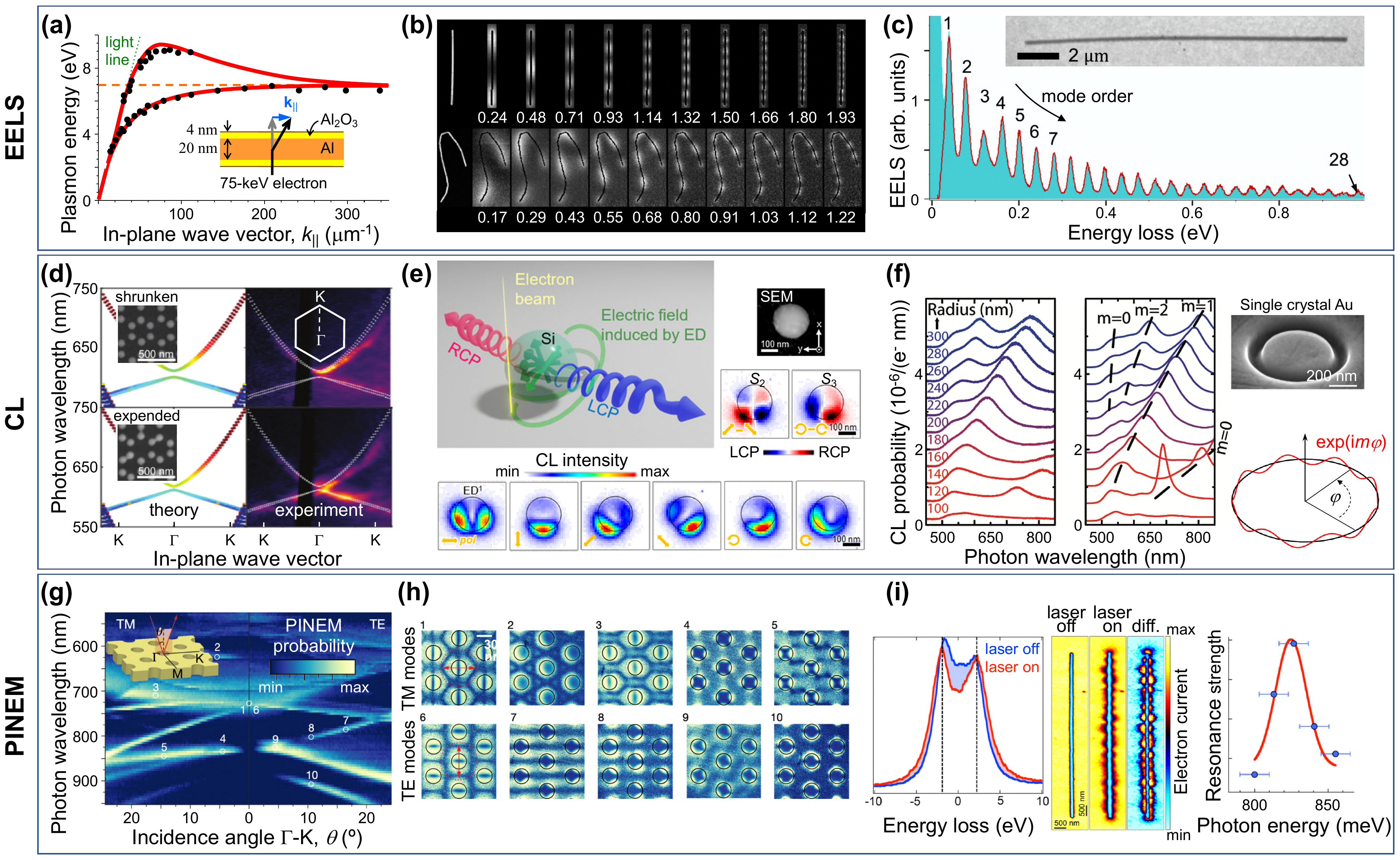}}
\caption{Probing nanoscale optical excitations. We show examples of mode dispersion relations (a,d,g), spatial mode distributions (b,e,h), and spectrally narrow plasmons (c,f,i) probed through EELS (a-c), CL (d-f), and PINEM (g-i).
(a) Plasmon dispersion measured in a self-standing aluminum film through angle- and energy-resolved transmitted electrons. Adapted from ref\ \citenum{PSV1975}. 
(b) Plasmon standing waves in long silver nanowires ($1.22\,\mu$m and $2.07\,\mu$m long in the top and bottom images, respectively) mapped by using 80\,keV TEM electrons and having energies (in eV) as indicated by labels. Adapted from ref\ \citenum{RB13}.
(c) Spectral features associated with high-quality-factor plasmon standing waves in a long copper nanowire ($15.2\,\mu$m length, 121\,nm diameter) extending from the mid- to the near-infrared, as resolved through high-resolution EELS. Adapted from ref\ \citenum{paperarxiv5}.
(d) Trivial and topological photonic crystal bands observed through 30\,keV SEM-based angle-resolved CL from two arrays of silicon pillars (200\,nm high, 88\,nm wide) deposited on a 10\,nm thick Si$_3$N$_4$ membrane and arranged on a hexagonal superlattice ($455$\,nm period) of either shrunken (138 hexagon side length) or expanded (168 side length) hexamers (see labels)  formed by six pillars per lattice site. Adapted from ref\ \citenum{PSN19}.
(e) Polarization-resolved CL intensity (lower maps) and emission Stokes parameters (center-right maps) produced by 80\,keV electrons in a TEM as a function of e-beam position over a silicon sphere (250\,nm diameter, see upper-right SEM image), as obtained by filtering $1.8\pm0.1\,$eV photons emitted with an angle of $45^\circ$ relative to the electron velocity. Adapted from ref\ \citenum{paperxx3}.
(f) Plasmon standing waves confined to circular grooves of different radii (see labels) carved into a single gold crystal (see upper-right SEM image) and mapped through CL, with the azimuthal number $m$ defining the number of periods along the circumference, as shown in the lower-right inset. Adapted from ref\ \citenum{paper137}.
(g,h) Dispersion relation (g) and near-field maps (h) of TM and TE modes in a 2D 200\,nm thick Si$_3$N$_4$ photonic crystal formed by a hexagonal hole array of 600\,nm period, mapped through PINEM using 80\,keV electrons. Adapted from ref\ \citenum{WDS20}.
(i) Silver nanowire plasmon standing wave spectrally resolved with 20\,meV accuracy (right) through the depletion observed in the zero-loss peak (ZLP) (left) as the frequency of the PINEM laser is scanned over the mode resonance. Adapted from ref\ \citenum{paper306}.}
\label{Fig1}
\end{figure*}

\begin{figure*}
\centering{\includegraphics[width=1.00\textwidth]{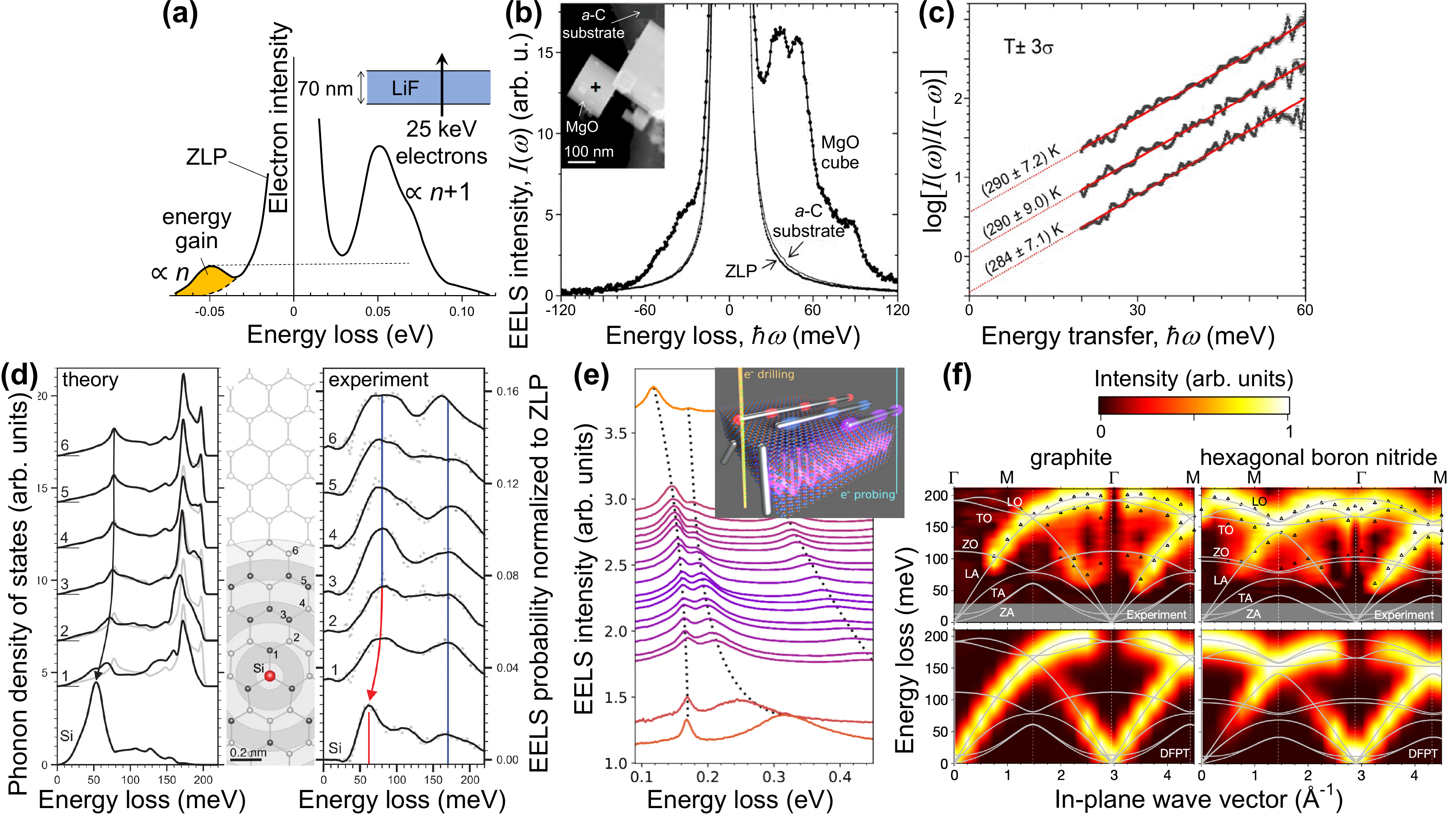}}
\caption{Electron-beam vibrational spectromicroscopy.
(a) Spectral features of phonon polaritons in LiF recorded through energy losses and gains experienced by 25\,keV electrons transmitted through a thin foil, with the gains originating in thermally populated modes at room temperature $T\approx300\,$K and the loss-to-gain peak ratio approximately given by $1+1/n_T(\omega)=\ee^{\hbar\omega/\kB T}$ ($\sim7$ at $\hbar\omega=50\,$meV). Adapted from ref\ \citenum{BGS1966}.
(b,c) Nanoscale e-beam thermometry based on high-resolution EELS of a MgO cube (b), whereby the sample temperature is determined upon examination of the loss-to-gain intensity ratio (c). Adapted from ref\ \citenum{LB18}.
(d) Atomic resolution in the mapping of vibrational spectra, here used to image the localization of the phonon density of states produced by a Si defect in monolayer graphene. Adapted from ref\ \citenum{HRK20}.
(e) Strong coupling between hBN photon polaritons and silver nanowire plasmons observed through high-resolution EELS by iterative e-beam drilling to shrink the wire length and scan one of its plasmon resonances over the phononic spectral region. Adapted from ref\ \citenum{paper342}.
(f) Phonon dispersion in graphite and hBN obtained by high-resolution angle-resolved EELS. Adapted from ref\ \citenum{SSB19}.}
\label{Fig2}
\end{figure*}

Spatial resolution down to the atomic scale can be achieved by using electrons as either probes or drivers of the sampled optical excitations. In particular, inelastically scattered beam electrons carry information on the excited states of the specimen, which can be revealed by performing electron energy-loss spectroscopy (EELS) \cite{E96,E03,EB05,B06}, as extensively demonstrated in the spectral and spatial mapping of optical modes covering a broad frequency range, stretching from the ultraviolet to the far infrared \cite{paper149,KS14,paper338,KUB09,KLD14,LTH17,LB18,HNY18,HHP19,HKR19,paper342,HRK20,YLG21,paper359}. Several examples of application are reviewed in Figures\ \ref{Fig1}a-c and \ref{Fig2}. In this field, benefiting from recent advances in instrumentation \cite{BDK02,KLD14,KDH19}, state-of-the-art transmission electron microscopes (TEMs) operated at $\sim30-300$\,kV acceleration voltages can currently deliver spectrally filtered images with combined sub-{\AA} and few-meV space-energy resolution \cite{KLD14,LTH17,LB18,HNY18,HHP19,HKR19,paper342,HRK20,YLG21,paper359} (see Figures\ \ref{Fig1}c and \ref{Fig2}d,e). Indeed, the reduction in the width of the electron zero-loss peak (ZLP) below $\sim10\,$meV and the ensuing high spectral resolution in EELS enable the exploration of optical modes down to the mid-infrared, including phonons in graphene \cite{HRK20} and silicon carbide \cite{YLG21} along with their modification due to atomic-scale defects  (Figure\ \ref{Fig2}d), phonons and phonon polaritons in graphite \cite{SSB19} and hexagonal boron nitride \cite{paper342,SSB19} (hBN) (Figure\ \ref{Fig2}e,f), and low-energy plasmons in long silver \cite{RB13} (Figure\ \ref{Fig1}b) and copper \cite{paperarxiv5} (Figure\ \ref{Fig1}c) nanowires. In addition, under parallel e-beam illumination, the inelastic electron signal can be resolved in energy and deflection angle to provide dispersion diagrams of surface modes in planar structures \cite{BGI1966,PSV1975,CS1975,CS1975_2,SSB19} (see Figures\ \ref{Fig1}a and \ref{Fig2}f). A vibrant field of e-beam vibrational spectromicroscopy has emerged in this context (see Figure\ \ref{Fig2}), with achievements such as the determination of the sample temperature distribution with nanometer precision thanks to the analysis of energy gains produced in the electrons by absorption of thermally populated modes \cite{LTH17,ILT18,LB18,paperarxiv5} (Figure\ \ref{Fig2}b,c), thus adding high spatial resolution to previous demonstrations of this approach \cite{BGS1966} (Figure\ \ref{Fig2}a).

A limiting factor in TEMs is imposed by the requirement of electron-transparent specimens with a total thickness of $\lesssim100\,$nm. At the cost of reducing spatial resolution, low-energy ($\sim50-500\,$eV) electron microscopy (LEEM) allows studying thicker samples by recording surface-reflected electrons \cite{R95}. This approach enables the acquisition of dispersion diagrams in planar surfaces by resolving the electron deflections associated with in-plane momentum transfers \cite{NHH01}, even in challenging systems such as monoatomic rows of gold atoms arranged on a vicinal silicon surface, which were neatly shown to support 1D plasmons through LEEM \cite{NYI06}. Likewise, using intermediate e-beam energies ($\sim1-50\,$keV), secondary electron microscopes (SEMs) offer the possibility of studying optical modes also in thick samples through the cathodoluminescence (CL) photon emission associated with the radiative decay of some of the created excitations \cite{paper149}, as extensively demonstrated in the characterization of localized \cite{paper035,paper116,paper137,paper167,WLC18} and propagating \cite{BJK06,VVP06,YS08} surface plasmons (see an example in Figure\ \ref{Fig1}f), as well as optical modes in dielectric cavities \cite{SCR12,paper341,paperxx3} (see Figure\ \ref{Fig1}e) and topological 2D photonic crystals \cite{PSN19} (see Figure\ \ref{Fig1}d), with spatial resolution in the few-nm range \cite{SMS19}. Some of these and other related studies were performed in TEMs \cite{YST96,paper035,KZ17,paper251,paper341,paperxx3}, where a direct comparison between CL and EELS was found to reveal similarities of the resulting spectra and those associated with optical elastic scattering and extinction, respectively \cite{paper251}. Combined with time-resolved detection, CL permits determining the lifetime and autocorrelation of sample excitations created by the probing electrons \cite{MSC05,TK13,MTC15,BMT16,MCW18,SMC20}, while the analysis of the angular distribution of the light emission provides direct information on mode symmetries \cite{paper116,SCR12,SAG20,paper341,paperxx3}. Nevertheless, EELS has the unique advantage of being able to detect dark optical excitations that do not couple to propagating radiation ({\it e.g.}, dark plasmons) but can still interact with the evanescent field of the passing electron probe \cite{KBK09,paper121,SDH12,BRF14}. In this respect, the presence of a substrate can affect the modes sampled in a nanostructure, for example by changing their optical selection rules, therefore modifying the radiation characteristics that are observed through CL \cite{SAG20,paper341}. Additionally, by collecting spectra for different orientations of the sample relative to the e-beam, both EELS \cite{NPL13} and CL \cite{ABC15} have been used to produce tomographic reconstructions of plasmonic near fields.

\begin{figure*}
\centering{\includegraphics[width=0.70\textwidth]{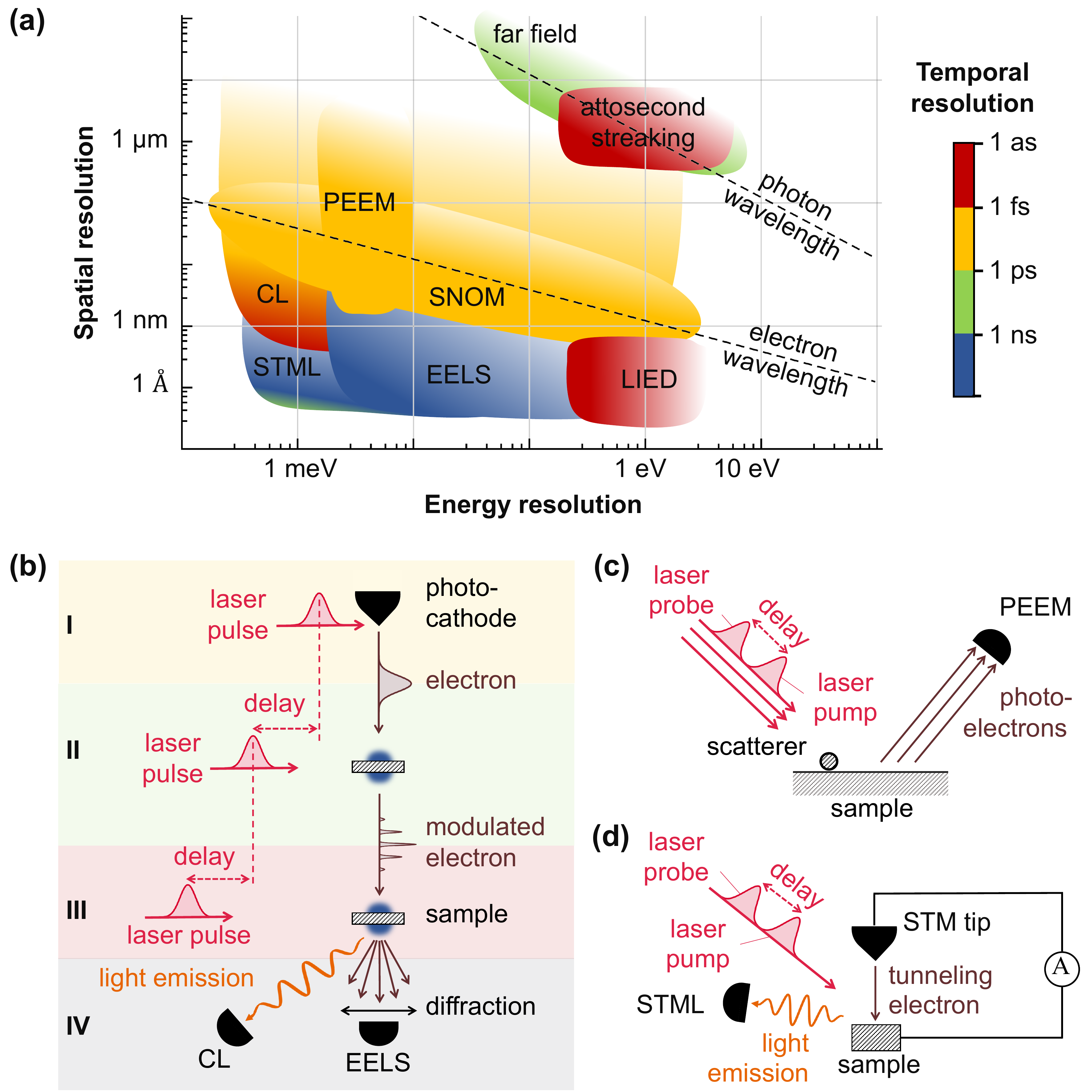}}
\caption{Microscopies at the frontier of space-time-energy resolution. (a) We organize different microscopy techniques according to their spatial (vertical axis), spectral (horizontal axis), and temporal (color scale) resolutions. The latter is limited to the sub-ns regime when relying on fast electronics \cite{MSC05} (green and blue), while it reaches the fs domain with optical pulses (yellow) and the attosecond range with X-ray pulses (red), but also with ultrashort electron pulses. In particular, measurement of CL driven by temporally compressed e-beams could potentially provide simultaneous sub-{\AA}--attosecond--sub-meV resolution (see main text). (b) Schematic illustration of an ultimate ultrafast electron microscope, encompassing (1) a photocathode tip that acts as an electron source driven by photoemission upon laser pulse irradiation; (2) an electron-modulation block based on PINEM-like interaction and subsequent free-space propagation that generates attosecond electron pulses; (3) a sample stage accessed by synchronized electron and laser pulses; and (4) the acquisition of several types of signals that include angle-resolved EELS and CL. The three fs laser pulses illuminating the photocathode, the sample, and the PINEM intermediate element are synchronized with attosecond-controlled delays. Currently available TEM and SEM setups incorporate different partial combinations of these possibilities. (c) Schematic illustration of time-resolved PEEM, where photoelectrons are used to construct fs- and nm-resolved movies by scanning the time delay between pump and probe laser pulses. (d) Illustration of STML, which enables atomic resolution through the detection of luminescence produced by inelastically tunneling electrons (right) and could be acquired with sub-ps temporal precision through modulation of the tip gate voltage. Femtosecond resolution could be potentially achieved through measurement of the laser-assisted electron tunneling current using pump-probe optical pulses (left).}
\label{Fig3}
\end{figure*}

The emergence of ultrafast transmission electron microscopy (UTEM) has added femtosecond (fs) temporal resolution to the suite of appealing capabilities of e-beams \cite{GLW06,BPK08,BFZ09,ARM20}. In this field, fs laser pulses are split into a component that irradiates a photocathode to generate individual fs electron pulses and another component that illuminates the sample with a well-controlled delay relative to the time of arrival of each electron pulse \cite{GLW06,BPK08,BFZ09} (Figure\ \ref{Fig3}b). Slow (sub-ps) structural changes produced by optical pumping have been tracked in this way \cite{GLW06,BPK08}, while the optical-pump--electron-probe (OPEP) approach holds the additional potential to resolve ultrafast electron dynamics \cite{HBL16,paperarxiv1}. It should be noted that an alternative method in UTEM, consisting in blanking the e-beam with sub-ns precision, can be incorporated in high-end SEMs and TEMs without affecting the beam quality \cite{paper325}, although with smaller temporal precision than the photocathode-based technique.

The electron-sample interaction is generally weak at the high kinetic energies commonly employed in electron microscopes, and consequently, the probability for an electron to produce a valence excitation or give rise to the emission of one photon is typically small ($\lesssim10^{-4}$). Nevertheless, low-energy electrons such as those used in LEEMs (and also in SEMs operated below $\sim1\,$keV) can excite individual nanoscale confined modes with order-unity efficiency \cite{paper228}, although a yield $\ll1$ should be expected in general at higher electron energies. The OPEP approach thus addresses nonlinear processes triggered by optical pumping and sampled in a perturbative ({\it i.e.}, linear) fashion by the electron \cite{GLW06,BPK08,paperarxiv1}. Furthermore, UTEM setups can produce multiple photon exchanges with each beam electron even if the specimen responds linearly to the optical pulse. Indeed, while a net absorption or emission of photons by the electron is kinematically forbidden in free space \cite{paper311}, the presence of the sample introduces evanescent optical field components that break the energy-momentum mismatch, leading to a nonvanishing electron-photon interaction probability, which is amplified by stimulated processes in proportion to the large number of incident photons ($\propto$ laser intensity) contained in each optical pulse. This effect has been argued to enable high spectral resolution by performing electron energy-gain spectroscopy (EEGS) while scanning the pumping light frequency \cite{H99,paper114,H09,paper306}, so that energy resolution is inherited from the spectral width of the laser, whereas the atomic spatial resolution of TEM setups can be retained. A similar approach has been followed to push energy resolution down to the few-meV range by analyzing the depletion of the ZLP upon intense laser irradiation \cite{paper306} (see Figure\ \ref{Fig1}i). We reiterate that potential the degradation of beam quality and energy width introduced at the photocathode can be avoided by resorting instead to e-beam blanking in combination with synchronized nanosecond laser pulses \cite{paper325}.

\begin{figure*}
\centering{\includegraphics[width=1.00\textwidth]{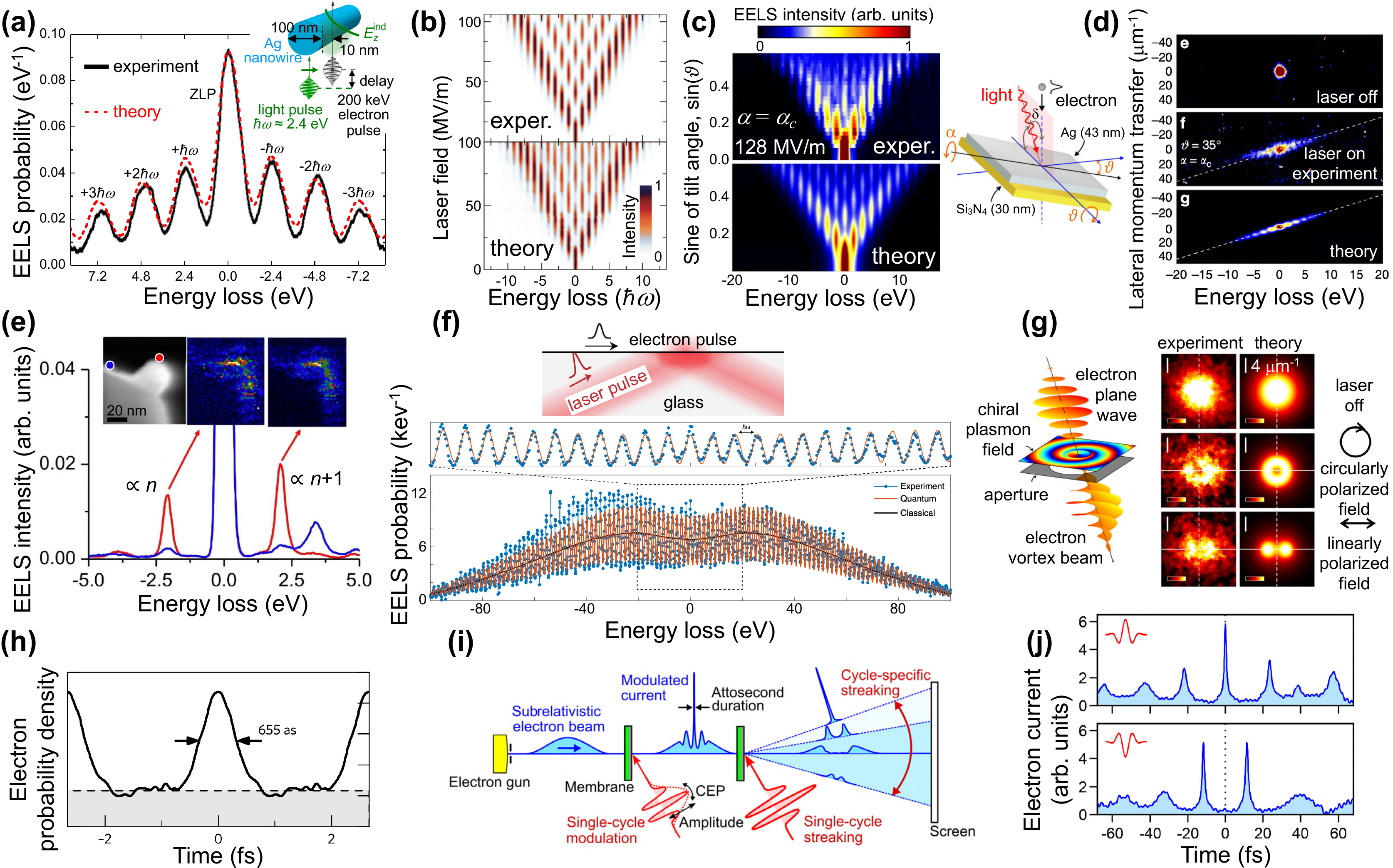}}
\caption{Optical modulation of free electrons.
(a) Energy comb of electron losses and gains produced by ultrafast interaction with evanescent light fields in the PINEM approach: experiment \cite{BFZ09} and theory \cite{paper151} comparison. Adapted from ref\ \citenum{paper151}.
(b) Laser-amplitude dependence of the electron energy comb produced by PINEM interaction, revealing quantum billiard dynamics among different electron energy channels separated by the photon energy $\hbar\omega$. Adapted from ref\ \citenum{FES15}.
(c,d) Tilt-angle dependence of the PINEM energy comb produced by using a planar film (c) and associated transfers of lateral linear momentum (d). Adapted from ref\ \citenum{paper311}.
(e) PINEM in the intermediate-coupling regime showing a $(n+1)/n$ loss-gain intensity ratio in the EELS spectra of silver nanoparticles with 100\,keV electrons under ns-laser illumination, superimposed on regular spontaneous EELS features, for beam positions as shown in the color-coordinated spots of the upper-left image, along with gain and loss energy-filtered images in the upper-middle and -right plots. Adapted from ref\ \citenum{paper325}.
(f) Intense-coupling regime resulting in a large number of PINEM energy sidebands under total-internal-reflection phase-matched illumination ({\it i.e.}, with the electron velocity matching the surface-projected light speed inside the glass). Adapted from ref\ \citenum{DNS20}.
(g) Transfer of angular momentum between light and electrons, as revealed in a configuration similar to (c) through a donut shape of the electron intensity in the Fourier plane after PINEM interaction. Adapted from ref\ \citenum{paper332}.
(h) Electron modulation into a train of attosecond pulses upon propagation from the PINEM interaction region over a sufficiently large distance to interlace different energy  sideband components in an electron microscope. Adapted from ref\ \citenum{PRY17}.
(i,j) Single electron pulses produced by streaking a train of pulses following the scheme shown in panel (i) and experimental demonstration based on the observation of the time-resolved electron current in a table-top e-beam-line setup (j). Adapted from ref\ \citenum{MB20}.
}
\label{Fig4}
\end{figure*}

In this context, intense efforts have been devoted to studying nonlinear interactions from the electron viewpoint in UTEM setups, assisted by the linear response of the sample to optical laser pumping. As a manifestation of these interactions, multiple quanta can be exchanged between the light and electron pulses in what has been termed photon-induced near-field electron microscopy (PINEM) \cite{BFZ09,paper151,PLZ10,PZ12,KGK14,PLQ15,FES15,paper282,EFS16,KSE16,RB16,VFZ16,paper272,PRY17,KML17,FBR17,paper306,paper311,paper312,MB18,MB18_2,paper325,paper332,K19,PZG19,paper339,RML20,DNS20,KLS20,WDS20,RK20,MVG20,paper360,VMC20}. The longitudinal (along the e-beam direction) free-electron wave function is then multiplexed in a periodic energy comb formed by sidebands separated from the ZLP by multiples of the laser photon energy \cite{BFZ09,paper151,PLZ10,PLQ15,FES15,EFS16} and associated with discrete numbers of net photon exchanges (Figure\ \ref{Fig4}a,b,c), the probability of which can be expressed in terms of a single coupling parameter $\beta$ that encapsulates the electron interaction with the optical near field and depends on lateral position in the transverse e-beam plane (see below). Such transverse dependence can be engineered to imprint an on-demand phase pattern on the electron wave function, giving rise, for example, to discretized exchanges of lateral linear momentum \cite{paper272,paper311,FYS20} (see Figure\ \ref{Fig4}d and also ref\ \citenum{FYS20} for sharper features associated with momentum discretization) and orbital angular momentum \cite{paper332,paper312} (Figure\ \ref{Fig4}g) between the light and the electron. PINEM spectral features ({\it i.e.}, the noted energy comb) do not bear phase coherence relative to spontaneous excitations associated with EELS \cite{paper325}, as experimentally verified for relatively low laser intensities, which lead to stimulated (PINEM loss and gain peaks) and spontaneous (EELS, only loss) energy peaks in the observed spectra with comparable strengths (Figure\ \ref{Fig4}e). In this regime, single-loss and -gain peak intensities are proportional to $n+1$ and $n$, respectively, where $n$ is the population of the laser-excited sample mode to which the electron couples. In contrast, we have $n\gg1$ at high laser fluence, so gain and loss features configure a symmetric spectrum with respect to the ZLP. As the intensity increases (Figure\ \ref{Fig4}a,b), multiple photon exchanges take place. These events were predicted \cite{paper151}, and subsequently confirmed in experiment \cite{FES15}, to give rise to a sub-fs quantum billiard dynamics (Figure\ \ref{Fig4}b). Enhanced order-unity electron-photon coupling is achieved under phase-matching conditions when the electron travels at the same velocity as the optical mode to which it couples \cite{paper180,K19}. Under this condition, the number of PINEM energy sidebands is strongly enlarged \cite{KLS20,DNS20} (see Figure\ \ref{Fig4}f), eventually reducing the loss-gain spectral symmetry, presumably due to departures from phase-matching produced by electron recoil. Incidentally, inelastic ponderomotive interactions can also be a source of asymmetry, as we discuss below, and so are corrections due to electron recoil \cite{T20}.

The optical near-field dynamics in nanostructures has been explored through PINEM, as illustrated by the acquisition of fs-resolved movies of surface plasmons evolving in nanowires \cite{PLQ15} and buried interfaces \cite{paper282}, as well as in the characterization of optical dielectric cavities and the lifetime of the supported optical modes \cite{KLS20,WDS20} (see Figure\ \ref{Fig1}g,h). It should be noted that analogous plasmon movies can be obtained through optical pump-probing combined with photoemission electron microscopy (PEEM, Figure\ \ref{Fig3}c) performed on clean surfaces \cite{MYH20}, as demonstrated for propagating plane-wave \cite{KOP05,KPP07}, chiral \cite{SKM17,DJD20}, and topological \cite{DZG20} plasmons. Nevertheless, by employing different types of particles to pump and probe ({\it e.g.}, photons and electrons), PINEM-modulated e-beams can potentially enable access into the attosecond regime without compromising energy resolution, as we argue below.

Complementing the above advances, the generation of temporally compressed electron pulses has emerged as a fertile research area \cite{BZ07,SCI08,PRY17,MB18_2,KES17,KSH18,MB18,SMY19,RTN20} that holds potential to push time resolution toward the attosecond regime. An initial proposal relied on free-space electron-light interactions \cite{BZ07}. Indeed, electron energy combs can also be produced in free space through ponderomotive interaction with two suitably oriented light beams of different frequencies $\omega_1$ and $\omega_2$ as a result of stimulated Compton scattering, subject to the condition $\omega_1-\omega_2=(\kb_1-\kb_2)\cdot\vb$, where $\kb_1$ and $\kb_2$ denote the photon wave vectors and $\vb$ is the electron velocity. The resulting electron spectrum consists of periodically spaced energy sidebands separated from the ZLP by multiples of the photon energy difference $\hbar|\omega_1-\omega_2|$ \cite{KES17}. After a long propagation distance beyond the electron-photon interaction region, different energy components in the electron wave function, traveling at slightly different velocities, become interlaced and can give rise to a periodic train of compressed-probability-density pulses with a temporal period $2\pi/|\omega_1-\omega_2|$. For sufficiently intense light fields, these pulses were argued to reach sub-fs duration \cite{BZ07}, as neatly confirmed in free-space experiments \cite{KES17,KSH18}. In a separate development, compression down to sub-fs pulses was achieved for spatially ($\sim100\,\mu$m) and spectrally ($\sim30\,$keV) broad multi-electron beams accelerated to 60\,MeV \cite{SCI08} using an inverse-free-electron-laser approach that relied on the coupling to the optical near field induced in a grating by irradiation with sub-ps laser pulses. In a tour-de-force experiment, PINEM-based production of attosecond pulse trains (Figure\ \ref{Fig4}h) was eventually pioneered in an electron microscope \cite{PRY17} at the single-electron level, yielding it compatible with $<1\,$nm e-beam spots and quasimonochromatic incident electrons ($<0.6\,$eV spread), thus raising the control over the electron wave function to an unprecedented level, and simultaneously rendering temporally modulated electrons accessible for use in spatially resolved spectroscopy. A demonstration of attosecond compression followed soon after using a table-top e-beam line setup \cite{MB18_2}, along with the generation of single electron pulses by subsequent angular sorting based on optical streaking \cite{MB20} (Figure\ \ref{Fig4}i,j), which is promising for the synthesis of individual attosecond electron pulses, although its combination with sub-nm lateral e-beam focusing in a microscope remains as a major challenge.

We organize the above-mentioned techniques in Figure\ \ref{Fig3}a according to their degree of space-time-energy resolution. Notably, electron-based methods offer better spatial resolution than all-optical approaches because of the shorter wavelength of such probes compared to photons. Incidentally, for the typical $30-300$\,keV e-beam energies, the electron wavelength lies in the $7-2$\,pm range, which sets an ultimate target for the achievable spatial resolution, currently limited by the numerical aperture of electron optics (NA$\sim10^{-2}$, leading to an e-beam focal size of $\sim0.5\,${\AA}). In contrast, far-field light optics and even SNOM offer a lower spatial resolution. We include for comparison laser-induced electron diffraction (LIED), which relies on photoemission from spatially oriented individual molecules produced by attosecond X-ray pulses, followed by electron acceleration driven by a synchronized infrared laser and subsequent elastic scattering back at the molecules; this technique grants us access into the molecular atomic structure with sub-{\AA}--attosecond precision \cite{WPL16} and it also provides indirect information of electronic potential-energy surfaces \cite{paper324}. Interestingly, time-resolved low-energy electron diffraction has also been employed to study structural dynamics in solid surfaces using photoemission e-beam sources analogous to UTEM \cite{VSH18}. In a radically different approach, scanning tunneling microscope luminescence \cite{KGM17} (STML, Figure\ \ref{Fig3}d) provides atomic spatial precision combined with optical spectral resolution in the determination of electronic defects in conducting surfaces \cite{LGD20,paper354}, which can in principle be combined with fast electronics to achieve sub-ns temporal resolution similar to CL \cite{MSC05}. Additionally, laser-driven tunneling in the STM configuration can provide fs resolution by measuring the electron current under optical pump-probe laser irradiation \cite{MPT02,DAZ11,KGM17} (Figure\ \ref{Fig3}c). In this article, we speculate that the team formed by synchronized ultrafast laser and free-electron pulses combined with measurement of angle-resolved CL (Figure\ \ref{Fig3}b) holds the potential to reach the sought-after sub-{\AA}--attosecond--sub-meV simultaneous level of resolution in the study of optical excitations, while even higher accuracy is still possible from the point of view of the fundamental limits (see below). These ideas can be implemented in TEMs, SEMs, and LEEMs, with the last two of these types of instruments presenting the advantage of offering stronger electron interaction with nanoscale optical modes.

\section{Fundamentals of Electron-Beam Spectroscopies}

Theoretical understanding of electron microscopy has benefited from a consolidated formalism for the analysis of EELS and CL spectra, as well as new emerging results in the field of UTEM. We present below a succinct summary of the key ingredients in these developments.

\subsection{Spontaneous Free-Electron Interaction with Sample Optical Modes} For the swift electron probes and low excitation energies under consideration, EELS and CL transition probabilities can be obtained by assimilating each beam electron to a point charge $-e$ moving with constant velocity vector $\vb=v\zz$ (nonrecoil approximation, see below) and interacting linearly with each sample mode. The electron thus acts as an external source of evanescent electromagnetic field, and in particular, the frequency decomposition of the electric field distribution as a function position $\rb=(\Rb,z)$ (with $\Rb=(x,y)$) for an electron passing by $\rb=(\Rb_0,0)$ at time zero admits the expression \cite{paper149}
\begin{align}
\Eb^{\rm ext}(\rb,\omega)=\frac{2e\omega}{v^2\gamma}\,\ee^{\ii\omega z/v}\,
\Fb(\Rb-\Rb_0,\omega),
\nonumber
\end{align}
where
\begin{align}
\Fb(\Rb,\omega)=\frac{\ii}{\gamma}K_0\left(\frac{\omega R}{v\gamma}\right)\zz-K_1\left(\frac{\omega R}{v\gamma}\right) \RR
\label{Fb}
\end{align}
and $\gamma=1/\sqrt{1-v^2/c^2}$ is the relativistic Lorentz factor. The time-dependent field is obtained through the Fourier transform
\[\Eb^{\rm ext}(\rb,t)=\frac{1}{2\pi}\int_{-\infty}^\infty\,d\omega\,\Eb^{\rm ext}(\rb,\omega)\,\ee^{-\ii\omega t}.\]
At large radial separations $R$, the two modified Bessel functions in $\Fb$ decay exponentially as $K_m(\zeta)\approx\ee^{-\zeta}\sqrt{\pi/2\zeta}$, whereas at short distances it is $K_1(\zeta)\approx1/\zeta$ that provides a dominant divergent contribution and explains the excellent spatial resolution of e-beams \cite{E07}. The induced field $\Eb^{\rm ind}$ acts back on the electron to produce a stopping force. By decomposing the resulting energy loss in frequency components, we can write the EELS probability as \cite{paper149}
\begin{align}
\Gamma_{\rm EELS}(\Rb_0,\omega)=\frac{e}{\pi\hbar\omega} \int_{-\infty}^\infty dz \,{\rm Re} \left\{\ee^{-\ii\omega z/v}E_z^{\rm ind}(\Rb_0,z,\omega) \right\}.
\label{EELS}
\end{align}
This quantity is normalized in such a way that $\int_0^\infty\,d\omega\,\Gamma_{\rm EELS}(\Rb_0,\omega)$ is the total loss probability and $\int_0^\infty\,d\omega\,\hbar\omega\,\Gamma_{\rm EELS}(\Rb_0,\omega)$ is the average energy loss.

It is convenient to express the EELS probability in terms of the $3\times3$ electromagnetic Green tensor $G(\rb,\rb',\omega)$, implicitly defined by the equation
\begin{align}
&\nabla \times \nabla \times G(\rb,\rb',\omega)-\frac{\omega^2}{c^2}\epsilon(\rb,\omega) G(\rb,\rb',\omega) \nonumber\\
&=-\frac{1}{c^2}\delta(\rb -\rb')
\label{Green}
\end{align}
for structures characterized by a local, frequency- and position-dependent permittivity $\epsilon(\rb,\omega)$ (and by an analogous relation for nonlocal media \cite{paper357}) and allowing us to obtain the induced field created by an external current $\jb^{\rm ext}(\rb,\omega)$ as
\[\Eb^{\rm ind}(\rb,\omega)=-4\pi\ii\omega\int d^3\rb'\,G^{\rm ind}(\rb,\rb',\omega)\cdot\jb^{\rm ext}(\rb',\omega).\]
The classical current associated with the electron is $\jb^{\rm ext}(\rb,\omega)=-e\,\zz\,\delta(\Rb-\Rb_0)\,\ee^{\ii\omega z/v}$, which upon insertion into the above expression, in combination with eq\ \ref{EELS}, yields
\begin{align}
\Gamma_{\rm EELS}(\Rb_0,\omega)=\frac{4e^2}{\hbar}&\int_{-\infty}^\infty dz\int_{-\infty}^\infty dz'\;\cos\left[\omega(z'-z)/v\right]\nonumber\\
&\times{\rm Im}\{-G_{zz}(\Rb_0,z,\Rb_0,z',\omega)\},
\label{EELSQM}
\end{align}
where we have replaced $G^{\rm ind}$ by $G$ because $G-G^{\rm ind}$ produces a vanishing contribution to the $z$ integrals as a consequence of kinematical mismatch between electrons and photons in free space \cite{paper149}. We remark the quantum nature of this result, which is revealed by the presence of $\hbar$, introduced through the lost energy $\hbar\omega$ in the denominator as a semiclassical prescription to convert the energy loss into a probability. This is also corroborated by a first-principles quantum-electrodynamics derivation of eq\ \ref{EELSQM}, which we offer in detail in the Appendix under the assumption that the sample is initially prepared at zero temperature.

An extension of this analysis to samples in thermal equilibrium at finite temperature $T$ allows us to relate the EELS probability to the zero-temperature result in eqs\ \ref{EELS} and \ref{EELSQM} as
\begin{align}
\Gamma_{\rm EELS}^T(\Rb_0,\omega)=\Gamma_{\rm EELS}(\Rb_0,|\omega|)\,\left[n_T(\omega)+1\right]\,{\rm sign}(\omega)
\label{EELST}
\end{align}
(with $\omega<0$ and $\omega>0$ indicating energy gain and loss, respectively), also derived in detail from first principles in the Appendix.

The far-field components of the induced field give rise to CL, with an emission probability that can be obtained from the radiated energy ({\it i.e.}, the time- and angle-integrated far-field Poynting vector). The classical field produced by the external electron source is thus naturally divided into frequency components, so an emission probability (photons per incident electron) is obtained by dividing by $\hbar\omega$, remarking again the quantum nature of the emission, which also reflects in how individual photon counts are recorded at the spectrometer in experiments. More precisely, using the external electron current and the Green tensor defined above, the electric field produced by the electron at a position $\rb_\infty$ far away from the sample can be written as
\begin{align}
\Eb^{\rm CL}(\rb_\infty,\omega)&=4\pi\ii e\omega \int_{-\infty}^\infty dz'\,\ee^{\ii\omega z'/v}\,G(\rb_\infty,\Rb_0,z',\omega)\cdot \zz \nonumber\\
&\xrightarrow[\omega r_\infty/c\to\infty]{} \frac{\ee^{\ii\omega r_\infty/c}}{r_\infty}\,\fb_{\rr_\infty}^{\rm CL}(\Rb_0,\omega),
\label{CLf}
\end{align}
where $\fb_{\rr_\infty}^{\rm CL}(\Rb_0,\omega)$ is the far-field amplitude. From the aforementioned analysis of the Poynting vector, we find that the CL emission probability reduces to
\[\Gamma_{\rm CL}=\int d^2\Omega_{\rr_\infty}\int_0^\infty d\omega\,\frac{d\Gamma_{\rm CL}}{d\Omega_{\rr_\infty}d\omega},\]
where \cite{paper149}
\begin{align}
\frac{d\Gamma_{\rm CL}}{d\Omega_{\rr_\infty}d\omega}=\frac{c}{4\pi^2\hbar\omega}\left|\fb_{\rr_\infty}^{\rm CL}(\Rb_0,\omega)\right|^2
\label{anothereq}
\end{align}
is the angle- and frequency-resolved probability.

A large number of EELS and CL experiments have been successfully explained using eq\ \ref{EELS} and the approach outlined above for CL by describing the materials in terms of their frequency-dependent local dielectric functions and finding $\Eb^{\rm ind}$ through numerical electromagnetic solvers, including the boundary-element method \cite{paper014,paper040,HK05,HDK09,HT12,paper197} (BEM) (see open-access implementation in ref\ \citenum{HT12}), the discrete-dipole approximation \cite{GH10,MGY12} (DDA), multiple scattering approaches \cite{paper025,TMH16}, and finite difference methods \cite{MNH10,CML15,DCP12}. Analytical expressions for the EELS and CL probabilities are also available for simple geometries, such as homogeneous planar surfaces, anisotropic films, spheres, cylinders, and combinations of these elements (see ref\ \citenum{paper149} for a review of analytical results), recently supplemented by an analysis of CL from a sphere for penetrating electron trajectories \cite{paperxx3}. It is instructive to examine the simple model of a sample that responds through an induced electric dipole, which admits the closed-form expressions
\begin{widetext}
\begin{align}
\left[\begin{matrix}
\Gamma_{\rm EELS}(\omega) \\
\Gamma_{\rm CL}(\omega)
\end{matrix}\right]
&=\frac{4e^2\omega^2}{\pi\hbar v^4\gamma^2}\;\times\left[\begin{matrix}
{\rm Im}\{\Fb^*(\Rb_0,\omega)\cdot\bar{\bar{\alpha}}(\omega)\cdot\Fb(\Rb_0,\omega)\} \\
(2\omega^3/3c^3)\left|\bar{\bar{\alpha}}(\omega)\cdot\Fb(\Rb_0,\omega)\right|^2
\end{matrix}\right] \nonumber\\
&=\frac{4e^2\omega^2}{\pi\hbar v^4\gamma^2}\;\left[K_1^2(\omega R_0/v\gamma)+\frac{1}{\gamma^2}K_0^2(\omega R_0/v\gamma)\right] \times\left[\begin{matrix}
{\rm Im}\{\alpha(\omega)\} \\
(2\omega^3/3c^3)\left|\alpha(\omega)\right|^2
\end{matrix}\right]
\label{EELSCLdip}
\end{align}
\end{widetext}
for the EELS and CL probabilities, where $\bar{\bar{\alpha}}(\omega)$ is the frequency-dependent $3\times3$ polarizability tensor, and the last equation applies to isotropic particles with $\bar{\bar{\alpha}}=\alpha(\omega)\mathcal{I}_3$. We remark that these results are quantitatively accurate even for large particles ({\it e.g.}, dielectric spheres sustaining Mie modes), provided we focus on spectrally isolated electric dipole modes \cite{paper149}. The above-mentioned properties of the $K_m$ functions readily reveal that the interaction strength diverges in the $R_0\rightarrow0$ limit ({\it i.e.}, when the e-beam intersects the point dipole). However, the finite physical sizes of the particle and the e-beam width prevent this divergence in practice. (Incidentally, the divergence also disappears in a quantum-mechanical treatment of the electron, which relates small $R_0$'s to large momentum transfers, limited to a finite cutoff imposed by kinematics.) In virtue of the optical theorem \cite{V1981} ({\it i.e.}, ${\rm Im}\{-1/\alpha(\omega)\}\ge2\omega^2/3c^3$), we have $\Gamma_{\rm EELS}\ge\Gamma_{\rm CL}$, as expected from the fact that emission events constitute a subset of all energy losses. Additionally, both EELS and CL share the same spatial dependence for dipolar modes, contained in the function $\Fb(\Rb_0,\omega)$ (eq\ \ref{Fb}).


As we show below, the transition probabilities are independent of the electron wave function, but a dependence is obtained in the partial electron inelastic signal when a selection is done on the incident and transmitted (final) wave functions ($\psi_i$ and $\psi_f$). Assuming a factorization of these wave functions as $\psi_{i|f}(\rb)\propto\psi_{i|f\perp}(\Rb)\,\ee^{\ii q_{i|f,z}z}/\sqrt{L}$, where $L$ is the quantization length along the beam direction, and integrating over longitudinal degrees of freedom (the $z$ coordinate), the state-selected transition probability depends on the transverse components as (see self-contained derivation in the Appendix)
\begin{widetext}
\begin{align}
\Gamma_{fi}(\omega)= \frac{4e^2}{\hbar} &\int d^2\Rb\int d^2\Rb' \;
\psi_{f\perp}(\Rb)\psi_{i\perp}^*(\Rb)\psi_{f_\perp}^*(\Rb')\psi_{i_\perp}(\Rb') \nonumber\\
&\times\int_{-\infty}^\infty dz\int_{-\infty}^\infty dz'\;\cos\left[\omega(z'-z)/v\right]\,{\rm Im}\left\{-G_{zz}(\rb,\rb',\omega)\right\}, \label{Gammafi}
\end{align}
\end{widetext}
where $G(\rb,\rb',\omega)$ is the electromagnetic Green tensor defined in eq\ \ref{Green}. Reassuringly, when summing $\Gamma_{fi}(\omega)$ over a complete basis set of plane waves for $\psi_{f\perp}(\Rb)$, we find $\sum_f\Gamma_{fi}(\omega)=\int d^2\Rb\,|\psi_{i\perp}(\Rb)|^2\,\Gamma_{\rm EELS}(\Rb,\omega)$, so we recover eq\ \ref{EELSQM} in the limit of a tightly focused incident beam ({\it i.e.}, $|\psi_{i\perp}(\Rb)|^2\approx\delta(\Rb-\Rb_0)$). Interestingly, the transition probability only depends on the product of transverse wave functions $\psi_{f_\perp}(\Rb)\psi_{i\perp}^*(\Rb)$. The possibility of selecting sample excitations by shaping this product has been experimentally confirmed by preparing the incident electron wave function in symmetric and antisymmetric combinations that excite dipolar or quadrupolar plasmons in a sample when the electrons are transmitted with vanishing lateral wave vector \cite{GBL17} ({\it i.e.}, for uniform $\psi_{f\perp}$ with $\qb_{f\perp}=0$). Similarly, under parallel beam illumination (uniform $\psi_{i\perp}$ with $\qb_{i\perp}=0$), angle-resolved Fourier plane imaging provides maps of transition probabilities to final states $\psi_{f\perp}\propto\ee^{\ii\qb_{f\perp}\cdot\Rb}$ of well-defined lateral momentum $\hbar\qb_{f\perp}$; actually, this approach is widely used to measure dispersion relations in planar films \cite{PSV1975,SSB19} (see Figures\ \ref{Fig1}a and \ref{Fig2}f), while a recent work tracks electron deflections produced by interaction with localized plasmons \cite{KGS18}. Analogously, the excitation of chiral sample modes by an incident electron plane wave produces vortices in the inelastically transmitted signal, an effect that has been proposed as a way to discriminate different enantiomers with nanoscale precision \cite{paper243}.

\subsection{Stimulated Free-Electron Interaction with Optical Fields} Under intense laser irradiation in UTEM setups, coupling to the optical near field in the sample region dominates the interaction with the electron. For typical conditions in electron microscopes, we can assume the electron to always consist of momentum components that are tightly focused around a central value $\qb_0$ parallel to the $z$ axis (nonrecoil approximation). This allows us to recast the Dirac equation into an effective Schr\"odinger equation \cite{PZ12},
\[(\partial_t+v\partial_z)\phi(\rb,t)=-\frac{\ii}{\hbar}\,\hat{\mathcal{H}}_1(\rb,t)\,\phi(\rb,t),\]
where we separate a slowly-varying envelope $\phi$ from the fast oscillations associated with the central energy $E_0$ and wave vector $\qb_0$ in the electron wave function \[\psi(\rb,t)=\ee^{\ii q_0z-\ii E_0t/\hbar}\phi(\rb,t)\] and we adopt the minimal-coupling light-electron interaction Hamiltonian \cite{paperarxiv4}
\[\hat{\mathcal{H}}_1=\frac{ev}{c}A_z+\frac{e^2}{2\me c^2\gamma}\left(A_x^2+A_y^2+\frac{1}{\gamma^2}A_z^2\right),\]
written in terms of the optical vector potential $\Ab(\rb,t)$ in a gauge with vanishing scalar potential without loss of generality. The nonrecoil approximation also implies that the initial electron wave function can be written as \[\psi_i(\rb,t)=\ee^{\ii q_0z-\ii E_0t/\hbar}\phi_i(\rb-\vb t),\] where $\phi_i$ defines a smooth invariant profile depending only on the rest-frame coordinates $\rb-\vb t$. Assuming that this behavior is maintained within the interaction region, the full electron wave function admits the solution \cite{T17}
\begin{align}
\psi(\rb,t)=\psi_i(\rb,t)\;\exp\left[\frac{-\ii}{\hbar}\int_{-\infty}^tdt'\,\hat{\mathcal{H}}_1(\rb-\vb t+\vb t',t')\right].
\nonumber
\end{align}
We focus for simplicity on monochromatic light of frequency $\omega$, for which the vector potential can be written as $\Ab(\rb,t)=(2c/\omega){\rm Im}\left\{\Eb(\rb)\ee^{-\ii\omega t}\right\}$, where $\Eb(\rb)$ is the optical electric field amplitude contributed by both the external laser and the components scattered by the sample. We are interested in evaluating the electron wave function at a long time after interaction, such that $\psi_i$ vanishes in the sample region. In this limit, combining the above results, we find that the transmitted wave function reduces to
\begin{align}
\psi(\rb,t)=&\psi_i(\rb,t)\;\ee^{\ii\varphi(\Rb)}\nonumber\\
&\times\mathcal{P}_0[\beta(\Rb),\omega,z-vt]\;\mathcal{P}_0[\beta'(\Rb),2\omega,z-vt],
\label{psiPINEM}
\end{align}
where
\begin{align}
\mathcal{P}_0(\beta,\omega,z)&=\exp\left(-\beta\ee^{\ii\omega z/v}+\beta^*\ee^{-\ii\omega z/v}\right) \nonumber\\
&=\sum_{l=-\infty}^\infty J_l(2|\beta|)\,\ee^{\ii l\,{\rm arg}\{-\beta\}}\,\ee^{\ii l\omega z/v}
\label{PPINEM}
\end{align}
describes the above-mentioned energy comb, associated with the absorption or emission of different numbers $l$ of photons of frequency $\omega$ by the electron, as ruled by the coupling coefficient
\begin{align}
\beta(\Rb)=\frac{e}{\hbar\omega}\int_{-\infty}^\infty dz\;E_z(\rb)\,\ee^{-\ii\omega z/v},
\label{beta}
\end{align}
which is determined by the optical field component along the e-beam direction. The rightmost expression in eq\ \ref{PPINEM} is derived by applying the Jacobi-Anger expansion $\ee^{\ii u\sin\theta}=\sum_lJ_l(u)\ee^{\ii l\theta}$ (eq\ 9.1.41 of ref\ \citenum{AS1972}) with $u=2|\beta|$ and $\theta={\rm arg}\{-\beta\}+\omega z/v$. The two other factors accompanying the incident wave function in eq\ \ref{psiPINEM} are produced by the ponderomotive force ({\it i.e.}, the $A^2$ term in the coupling Hamiltonian $\hat{\mathcal{H}}_1$). Namely, a phase
\begin{align}
\varphi(\Rb)=\frac{-1}{\mathcal{M}\omega^2}\int_{-\infty}^\infty dz\;\left[|E_x(\rb)|^2+|E_y(\rb)|^2+\frac{1}{\gamma^2}|E_z(\rb)|^2\right],
\label{phase}
\end{align}
where $\mathcal{M}=\me\gamma v/c\alpha$ plays the role of an effective mass and $\alpha\approx1/137$ is the fine structure constant; and an extra energy comb of double frequency given by eq\ \ref{PPINEM} with $\omega$ substituted by $2\omega$ and $\beta$ by
\begin{align}
\beta'(\Rb)=&\frac{-\ii}{2\mathcal{M}\omega^2} \label{betap}\\
&\times\int_{-\infty}^\infty dz\;\left[E_x^2(\rb)+E_y^2(\rb)+\frac{1}{\gamma^2}E_z^2(\rb)\right]\,\ee^{-2\ii\omega z/v}.
\nonumber
\end{align}
We remark that the multiplicative factors in eq\ \ref{psiPINEM} depend on the transverse coordinates $\Rb=(x,y)$. In the absence of a scattering structure, $\beta$ and $\beta'$ vanish, yielding $\mathcal{P}_0=1$ as a result of the aforementioned electron-photon kinematic mismatch, although a spatially modulated ponderomotive phase $\varphi$ can still be produced, for example by interfering two counter-propagating lasers, giving rise to electron diffraction (the Kapitza-Dirac effect \cite{KD1933,FAB01,B07,TL19}). From an applied viewpoint, this phenomenon enables optical sculpting of e-beams in free space \cite{MJD10,SAC19,ACS20,paperarxiv4}.

The relative strength of $A^2$ interactions can be estimated from the ratio $|\beta'/\beta|\sim |\Eb|/E_{\rm thres}$ (see eqs\ \ref{beta} and \ref{betap}), where $E_{\rm thres}=2\me\gamma v\omega/e$ ($\approx5\times10^{12}\,$V/m for $\hbar\omega=1.5\,$eV and 100\,keV electrons) defines a threshold field amplitude that exceeds by $\sim4$ orders of magnitude the typical values used in PINEM experiments \cite{FES15,paper311}, although they should be reachable using few-cycle laser pulses in combination with nonabsorbing high-index dielectric structures.

Neglecting $A^2$ corrections, the remaining PINEM factor trivially satisfies the relation $\mathcal{P}_0(\beta_1,\omega,z)\times\mathcal{P}_0(\beta_2,\omega,z)=\mathcal{P}_0(\beta_1+\beta_2,\omega,z)$ (see eq\ \ref{PPINEM}), so that the effect of two simultaneous or consecutive PINEM interactions with mutually coherent laser pulses at the same photon frequency is equivalent to a single one in which the coupling coefficient is the sum of the individual coupling coefficients, as neatly demonstrated in double-PINEM experiments \cite{EFS16}. Additionally, $\beta$ imprints a lateral dependent phase $l\,{\rm arg}\{-\beta(\Rb)\}$ on the wave function component associated with each inelastic electron sideband, where $l$ labels the net number of exchanged photons; this effect has been experimentally verified through the observation of transverse linear \cite{paper311,FYS20} and angular \cite{paper332} momentum transfers to the electron (Figure\ \ref{Fig4}d,g), and it has been predicted to produce electron diffraction by plasmon standing waves in analogy to the Kapitza-Dirac effect \cite{paper272}.

The Schr\"odinger equation mentioned at the beginning of this section neglects the effect of recoil, which can substantially affect the electron over long propagation distances $d$ beyond the PINEM interaction region. Incidentally, recoil can even manifest within the interaction region if it spans a relatively large path length. Neglecting again $A^2$ terms, the leading longitudinal recoil correction results in the addition of an $l$-dependent phase $-2\pi l^2d/z_T$ to each term of the sum in eq\ \ref{PPINEM}, where
\[z_T=\frac{4\pi\me v^3\gamma^3}{\hbar\omega^2}\]
is a Talbot distance ({\it e.g.}, $z_T\approx159$mm for $\hbar\omega=1.5\,$eV and 100\,keV electrons) that indeed increases with kinetic energy. More precisely, the electron wave function becomes $\psi(\rb,t)=\psi_i(\rb,t)\,\mathcal{P}_d[\beta(\Rb),\omega,z-vt]$, where
\begin{align}
\mathcal{P}_d(\beta,\omega,z)=\sum_{l=-\infty}^\infty J_l(2|\beta|)\,\ee^{\ii l\,{\rm arg}\{-\beta\}+\ii l\omega z/v-2\pi\ii l^2d/z_T}.
\label{PPINEMd}
\end{align}
We remark that this result is valid if we neglect ponderomotive forces and assume the e-beam to be sufficiently well collimated as to dismiss lateral expansion during propagation along the distance $d$. We also assume that $\psi_i$ is sufficiently monoenergetic as to dismiss its drift along $d$. Different $l$ components move with different velocities, resulting in a temporal compression of the electron wave function \cite{SCI08} that has been demonstrated to reach the attosecond regime \cite{KML17,PRY17,MB18_2,MB18,SMY19,RTN20,MB20}.

The above results refer to coherent laser illumination, but additional possibilities are opened by using quantum light instead, and in particular, we have predicted that the electron spectra resulting from PINEM interaction with optical fields carry direct information on the light statistics \cite{paper339} ({\it e.g.}, the second-order autocorrelation function $g^{(2)}$). Additionally, temporal electron pulse compression can be accelerated using phase-squeezed light (see Figure\ \ref{Fig7}d below), while the electron density matrix acquires nontrivial characteristics with potential application in customizing its eventual interaction with a sample \cite{paper360}.

The extension of the above results to multicolor illumination opens additional possibilities, with the linear $A$ term in $\hat{\mathcal{H}}_1$ producing multiplicative PINEM factors (one per light frequency) that lead to asymmetric electron spectra \cite{PRY17}. Also, the ponderomotive-force $A^2$ term introduces frequency-sum and frequency-difference PINEM factors, which in free space, with lasers arranged under phase-matching propagation directions, can give rise to energy combs similar to PINEM through stimulated Compton scattering \cite{KSH18}; this effect, combined with free-space propagation, has been exploited to achieve attosecond electron compression without requiring material coupling structures \cite{KES17}.

\subsection{Relation between PINEM and CL} In CL, the electron acts as a source from which energy is extracted to produce light emission, whereas PINEM is just the opposite: an external light source exchanges energy with the electron. It is thus plausible that a relation can be established between the two types of processes if the sample exhibits a reciprocal response, so that the electromagnetic Green tensor satisfies the property $G_{aa'}(\rb,\rb',\omega)=G_{a'a}(\rb',\rb,\omega)$, where $a$ and $a'$ denote Cartesian components. To explore this idea, we start from the PINEM coupling coefficient defined in eq\ \ref{beta} and consider far-field illumination from a well-defined direction $\rr_\infty$, as produced by an external distant dipole $\pb^{\rm ext}\perp\rr_\infty$ at the laser source position $\rb_\infty$. Using the Green tensor to relate this dipole to the electric field as $\Eb(\rb)=-4\pi\omega^2\,G(\rb,\rb_\infty,\omega)\cdot\pb^{\rm ext}$, we find 
\begin{align}
\beta(\Rb)=-\frac{4\pi e\omega}{\hbar}\sum_a p_a^{\rm ext} \int_{-\infty}^\infty dz\;G_{za}(\rb,\rb_\infty,\omega)\,\ee^{-\ii\omega z/v}.
\nonumber
\end{align}
In the absence of a sample, the external laser field is obtained from the far-field limit of the free-space Green tensor, giving rise to an external plane-wave of electric field $\Eb(\rb)=\Eb^{\rm ext}\ee^{\ii\kb\cdot\rb}$ with wave vector $\kb=-\rr_{\infty}\omega/c$ and amplitude $\Eb^{\rm ext}=\left(\ee^{\ii\omega r_\infty/c}/r_\infty\right)\,(\omega^2/c^2)\,\pb^{\rm ext}$, which allows us to recast the coupling coefficient into
\begin{align}
&\beta(\Rb)=\frac{\ii c^2}{\hbar\omega^2}\sum_a E_a^{\rm ext} \label{betaCL1}\\
&\times\left[4\pi\ii e\omega\frac{r_\infty}{\ee^{\ii\omega r_\infty/c}}\int_{-\infty}^\infty dz\;G_{az}(\rb_\infty,\rb,\omega)\,\ee^{-\ii\omega z/v}\right],
\nonumber
\end{align}
where we have used the noted reciprocity property. Now, we identify the expression inside square brackets as the CL far-field amplitude by comparison to eq\ \ref{CLf}. Finally, we find
\begin{align}
\beta(\Rb)=\frac{\ii c^2}{\hbar\omega^2}\;\tilde{\fb}_{\rr_\infty}^{\rm CL}(\Rb,\omega)\cdot\Eb^{\rm ext},
\label{betaCL2}
\end{align}
where the tilde in $\tilde{\fb}_{\rr_\infty}^{\rm CL}(\Rb_0,\omega)$ indicates that it has to be calculated for an electron moving with opposite velocity ({\it i.e.}, $-\vb$ instead of $\vb$; {\it cf.} $\ee^{\pm\ii\omega z/v}$ factors in eqs\ \ref{CLf} and \ref{betaCL1}). Equation\ \ref{betaCL2} establishes a direct relation between PINEM and CL: the coupling coefficient in the former, for far-field plane-wave illumination from a given direction $\rr_\infty$ ({\it i.e.}, light propagating toward $-\rr_\infty$), is readily obtained from the electric far-field amplitude of CL light emitted toward $\rr_\infty$, but with the electron velocity set to $-\vb$ instead of $\vb$. A recent study has partially verified this relation by exploring the spatial characteristics of EELS, CL, and PINEM for the same single gold nanostar \cite{paperarxiv7}. For completeness, we provide the expression
\[\beta(\Rb)=\frac{2\ii e\omega}{\hbar v^2\gamma}\;\alpha(\omega)\,\Fb^*(\Rb,\omega)\cdot\Eb^{\rm ext}\]
obtained for an isotropic dipolar scatterer (see eqs\ \ref{Fb} and \ref{EELSCLdip}) under continuous-wave illumination conditions.

The high degree of control over the free-electron wave function embodied by the above developments opens exciting opportunities to explore new physics and applications. However, before presenting some perspectives on these possibilities, we discuss in more detail the role of the electron wave function in the interaction with optical sample modes.

\section{Quantum and Classical Effects Associated with the Free-Electron Wave Function}

Like for any elementary particle, the wave nature of free electrons manifests in interference phenomena observed through double-slit experiments and diffraction by periodic lattices, which are typical configurations used to image material structures and their excitation modes. Electron interference has been extensively exploited in TEMs to this end \cite{HS1972,B94,E96,E03,EB05,B06,MD09,SKK17}, as well as in photoelectron diffraction \cite{F10_2}, low-energy electron diffraction \cite{P1974}, and LIED \cite{WPL16}. Shaping and probing the electron wave function lies at the heart of these techniques, in which the electrons are scattered elastically, and consequently, no final sample excitations are produced. Likewise, interference is expected to show up, associated with the creation of sample excitations by e-beams, as demonstrated in the so-called inelastic electron holography \cite{LF00,H08_2}.

It should be noted that electron beam spectroscopies involve the creation of excitations in the sample by one electron at a time when using typical beam currents $\lesssim1\,$nA ({\it i.e.}, $\lesssim6$ electrons per nanosecond). Such relatively low currents are employed to avoid Coulomb electron-electron repulsion and the resulting beam degradation and energy broadening, which are detrimental effects for spatially resolved EELS, although they can still be tolerated in diffraction experiments relying on electron bunches to retrieve structural information \cite{BHM20}, and also in EEGS based on depletion of the ZLP with few-meV energy resolution obtained by tuning the laser frequency \cite{paper306}. Understandably, the quantum character of individual electrons has been explored to pursue applications such as cavity-induced quantum entanglement \cite{K19,ZSF21}, qubit encoding \cite{RML20}, and single-photon generation \cite{paper180}.

Now, a recurrent question arises \cite{RH1977,paper149,GBL17,RKT19,PG19,GY20,paperarxiv3,paperarxiv6,paper360}, can the excitation efficiency be modulated by shaping the electron wave function? For single monoenergetic electrons, nonretarded theory was used to show that the excitation probability reduces to that produced by a classical point charge, averaged over the intensity of the transverse beam profile \cite{RH1977}. This result was later generalized to include retardation \cite{paper149}, and the predicted lack of dependence on transverse electron wave function was experimentally corroborated for Smith-Purcell radiation emission \cite{RKT19}. Some dependence can however be observed in EELS by collecting scattered electrons only within a partial angular range, as neatly demonstrated by Ritchie and Howie \cite{RH1977} in the nonretarded limit. This result was later generalized to include retardation \cite{paper149}. Specifically, for transmission along the center of the Fourier plane in an electron microscope, wave function shaping was experimentally demonstrated to actively select plasmon losses of dipolar or quadrupolar symmetry in metallic nanowires \cite{GBL17}.

The dependence on the longitudinal wave function is not as clear, and for example, a recent report \cite{GY20} based on a semiclassical description of the electric field generated by free electrons claims that the probability of exciting a sample initially prepared in the ground state could be enhanced for an individual electron distributed along a periodic density profile. However, this conclusion is inconsistent with a fully quantum-mechanical treatment of the electron-sample system (see detailed analysis below). Importantly, the same study claims that $N$ electrons arriving at random times produce an overall probability $\propto N^2$ when they are previously PINEM-modulated by the same laser, an effect that is indeed supported by a quantum description of the electrons, as we show below. In addition, a wave function dependence should be observed for interaction with samples prepared in a coherent superposition of ground and excited states that is phase-locked with respect to the electron wave function, as experimentally illustrated in double-PINEM experiments \cite{EFS16} (see below). While PINEM commonly relies on bosonic sample modes, an extension of this effect to two-level systems has also been discussed in recent theoretical works \cite{PG19,ZSF21}.

In this section, we elucidate the role of the electron wave function in the excitation of sample modes for any type of interactions with matter, photons, and polaritons. We derive analytical expressions from first-principles for the excitation probability produced by single and multiple electrons with arbitrarily shaped wave functions, based on which we conclude that the excitation by single electrons with the specimen prepared in any stationary state ({\it e.g.}, the ground state) can be described fully classically with the electron treated as a point particle, regardless of its wave function, apart from a trivial average over the transverse beam profile. In contrast, multiple electrons give rise to correlations between their respective wave functions, which enter through the electron probability densities, whereas phase information is completely erased. More precisely, the few-electron case (see analysis for two electrons below) reveals a clear departure from the classical point-particle picture, while in the limit of many electrons $N$, a classical description prevails, leading to an excitation probability $\propto N^2$ if they are bunched with a small temporal width relative to the optical period of the sampled excitation \cite{UGK98} or if their probability density is optically modulated with a common coherent light field \cite{NS1954,SH1969,UGK98,FFK1971,SCI08,GY20}. Crucially, these results follow from the nonrecoil approximation ({\it i.e.}, the fact that the electron velocity can be considered to be constant during the interaction), which accurately applies under common conditions in electron microscopy (small beam-energy spread and low excitation energies compared with the average electron energy). Our hope is that the present discussion clarifies current misunderstandings on the role of the electron wave function in inelastic scattering and provides simple intuitive rules to tackle configurations of practical interest.

\subsection{Lack of Wave-Function Dependence for a Single Electron} We first consider a free electron propagating in vacuum and interacting with arbitrarily shaped material structures. Without loss of generality, the wave function of this combined electron-sample system can be decomposed as
\begin{align}
|\psi(t)\rangle=\sum_{\qb n}\alpha_{\qb n}(t)\ee^{-\ii(\varepsilon_\qb+\omega_n)t}|\qb n\rangle
\label{defpsi}
\end{align}
using a complete basis set of combined material (and possibly radiation) states $|n\rangle$ of energy $\hbar\omega_n$ and electron plane-wave states $|\qb\rangle$ of well-defined momentum $\hbar\qb$ and energy $\hbar\varepsilon_\qb$. The elements of this basis set are eigenstates of the noninteracting Hamiltonian $\hat{\mathcal{H}}_0$, so they satisfy $\hat{\mathcal{H}}_0|\qb n\rangle=\hbar(\varepsilon_\qb+\omega_n)|\qb n\rangle$. This description is valid as long as no bound states of the electrons are involved. Under common conditions in electron microscopes, the states $|n\rangle$ describe excitations in the sample, including the emission of photons, but also undesired excitations in other parts of the microscope ({\it e.g.}, phonons in the electron source). For simplicity, we assume the electron to be prepared in a pure state $\sum_{\qb} \alpha^0_{\qb} |\qb \rangle$ and the sample in a stationary state $n=0$ prior to interaction ({\it i.e.}, $\alpha_{\qb n}(-\infty)=\delta_{n0}\alpha_\qb^0$, subject to the normalization condition $\sum_\qb|\alpha_\qb^0|^2=1$), in the understanding that the mentioned undesired excitations can later be accounted for by tracing over different incoherent realizations of the electron wave function in the beam.

By inserting eq\ \ref{defpsi} into the Schr\"odinger equation $(\hat{\mathcal{H}}_0+\hat{\mathcal{H}}_1)|\psi\rangle=\ii\hbar\partial_t|\psi\rangle$, where the Hamiltonian $\hat{\mathcal{H}}_1$ describes electron-sample interactions, we find the equation of motion for $n\neq0$
\[\ii\hbar\dot{\alpha}_{\qb n}=\sum_{\qb'n'}\ee^{\ii(\varepsilon_\qb-\varepsilon_{\qb'}+\omega_n-\omega_{n'})t}\langle\qb n|\hat{\mathcal{H}}_1|\qb'n'\rangle\alpha_{\qb'n'}\]
for the expansion coefficients $\alpha_{\qb n}$. Now, the results presented in this section are a consequence of the following two assumptions, which are well justified for typical excitations probed in electron microscopy \cite{paper149}:

(i) {\it Weak Coupling.} The electron interaction with the sample is sufficiently weak as to neglect higher-order corrections to the excitation probability beyond the first order. This allows us to rewrite the equation of motion as $\ii\hbar\dot{\alpha}_{\qb n}=\sum_{\qb'}\ee^{\ii(\varepsilon_\qb-\varepsilon_{\qb'}+\omega_{n0})t}\langle\qb n|\hat{\mathcal{H}}_1|\qb'0\rangle\alpha^0_{\qb'}$ (with $\omega_{n0}=\omega_n-\omega_0$), which can be integrated in time to yield the solution
\begin{align}
\alpha_{\qb n}(\infty)=-\frac{2\pi\ii}{\hbar}\sum_{\qb'}\delta(\varepsilon_\qb-\varepsilon_{\qb'}+\omega_{n0})\langle\qb n|\hat{\mathcal{H}}_1|\qb'0\rangle\alpha^0_{\qb'}
\label{alphainfinity}
\end{align}
for the wave function coefficients after interaction. We remark that $n=0$ can be the ground state or any excited state in the present derivation, as long as it is stationary.

(ii) {\it Nonrecoil Paraxial Approximation.} Electron beams feature small divergence angle ($\sim$ a few mrad) and low energy spread compared with the mean electron energy ({\it i.e.}, $\alpha_{\qb n}$ is negligible unless $|\qb-\qb_0|\ll q_0$, where $\hbar\qb_0$ is the central electron momentum). Additionally, we assume that the interaction with the sample produces wave vector components also satisfying $|\qb-\qb_0|\ll q_0$. This allows us to write the electron frequency difference as
\begin{align}
\varepsilon_\qb-\varepsilon_{\qb'}\approx\vb\cdot(\qb-\qb'),
\label{nonrecoil}
\end{align}
indicating that only momentum transfers parallel to the beam contribute to transfer energy to the sample \cite{paper149}. The nonrecoil approximation is generally applicable in the context of electron microscopy, unless the excitation energy is a sizeable fraction of the electron kinetic energy \cite{T20,WRM21}.

Putting these elements together and using the real-space representation of the electron states $\langle\rb|\qb\rangle=V^{-1/2}\,\ee^{\ii\qb\cdot\rb}$ with quantization volume $V$ in eq\ \ref{alphainfinity}, we find that the probability that a single beam electron excites a sample mode $n$, expressed through the trace of scattered electron degrees of freedom $\Gamma_n^0=\sum_\qb|\alpha_{\qb n}(\infty)|^2$, reduces to (see Appendix)
\begin{align}
\Gamma_n^0=\int d^3\rb\;|\psi^0(\rb)|^2 \,|\tilde{\beta}_n(\Rb)|^2
\label{P0n}
\end{align}
where
\begin{align}
\psi^0(\rb)=V^{1/2}\int\frac{d^3\qb}{(2\pi)^3}\,\alpha^0_\qb\,\ee^{\ii\qb\cdot\rb}
\label{psi0}
\end{align}
is the incident electron wave function,
\begin{align}
\tilde{\beta}_n(\Rb)=\frac{1}{\hbar v}\int_{-\infty}^\infty dz\;\ee^{-\ii\omega_{n0}z/v}\langle0|\hat{\mathcal{H}}_1(\rb)|n\rangle
\label{betan}
\end{align}
is an electron-sample coupling coefficient that depends on the transverse coordinates $\Rb=(x,y)$, and we choose the beam direction along $\zz$. We note that this definition of $\tilde{\beta}_n$ coincides with previous studies in which $\hat{\mathcal{H}}_1$ describes electron-light PINEM interaction and $n$ refers to optical modes \cite{paper339,paper360}. Also, the PINEM coupling coefficient in eq\ 11 is obtained from eq\ 22 by multiplying it by the laser-driven amplitude associated with mode $n$ and summing over $n$.

We observe from eq\ \ref{P0n} that the excitation probability does not depend on the electron wave function profile along the beam direction $\zz$, because this enters just through an integral of the electron density along that direction. Additionally, the dependence on transverse directions $\Rb$ consists of a weighted average of the probability $|\tilde{\beta}_n(\Rb)|^2$ over the transverse profile of the beam intensity.

\subsection{Wave-Function Dependence in the Correlation Among Multiple Electrons} The above analysis can readily be extended to a beam bunch consisting of $N$ distinguishable electrons with incident wave functions $\psi^j(\rb)$ labeled by $j=0,\dots,N-1$. The probability of exciting a sample mode $n$ then reduces to (see detailed derivation in the Appendix)
\begin{align}
\Gamma_n^{\rm total}=\sum_j\int d^3\rb\;|\psi^j(\rb)|^2 \,|\tilde{\beta}_n(\Rb)|^2+\sum_{j\neq j'} Q_n^jQ_n^{j'*}, \label{PNn}
\end{align}
where
\begin{align}
&Q_n^j=\int d^2\Rb \; M_n^j(\Rb)\tilde{\beta}_n(\Rb), \label{Qj}\\
&M_n^j(\Rb)=\int_{-\infty}^\infty dz\;\ee^{\ii\omega_{n0}z/v}\,|\psi^j(\rb)|^2. \label{Mnj}
\end{align}
The first term in eq\ \ref{PNn} corresponds to the sum of uncorrelated excitation probabilities produced by $N$ independent electrons, each of them expressed as a weighted average over the transverse electron density profile, just like for a single electron in eq\ \ref{P0n}. The second term accounts for two-electron correlations, in which the phase of the electron wave functions is also erased, but there is however a dependence on the electron probability densities through their Fourier transforms in eq\ \ref{Mnj}. Interestingly, the factor $|M_n^j(\Rb)|^2$ is in agreement with the result obtained for excitation with a classical charge distribution having the same profile as the electron probability density, which is well studied in the context of beam physics \cite{NS1954,SCI08}. Also, this factor has recently been identified as a measure of the degree of coherence of the electron in its interaction with mutually phase-locked external light \cite{paperarxiv3,paperarxiv6}. Obviously, $|M_n^j(\Rb)|^2$ is bound by the inequality $|\int d^2\Rb\, M_n^j(\Rb)|\le1$, with the equal sign standing for any value of the excitation frequency $\omega_{n0}$ in the limit of point-particle electrons ({\it i.e.}, $|\psi^j(\rb)|^2=\delta(\rb-\rb_j)$), and also for a fixed $\omega_{n0}$ and its multiples if the electron probability density is periodically modulated as
\begin{align}
|\psi^j(\rb)|^2=|\psi^j_\perp(\Rb)|^2\,\sum_s b_{j,s}\,\delta\left(z-z_0-\frac{2\pi sv}{\omega_{n0}}\right)
\label{psicomb}
\end{align}
with arbitrary coefficients $b_{j,s}$ ({\it i.e.}, a train of temporally compressed pulses separated by a spatial period $v/\omega_{n0}$). Periodically modulated electrons with a limited degree of compression are currently feasible through strong PINEM interaction followed by free-space propagation.

In the derivation of these results, we have assumed electrons prepared in pure states ({\it i.e.}, with well-defined wave functions). The extension to mixed electron states requires dealing with the joint electrons-sample density matrix elements $\rho_{\{\qb\} n,\{\qb'\}n'}(t)$ and calculating $\Gamma_n^{\rm total}=\sum_{\{\qb\}}\rho_{\{\qb\} n,\{\qb\}n}(\infty)$. Starting with $\rho_{\{\qb\} n,\{\qb'\}n'}(-\infty)=\delta_{n0}\delta_{n'0}\prod_j\rho^j_{\qb_j\qb'_j}$, where $\rho^j_{\qb_j\qb'_j}$ are the matrix elements of electron $j$ before interaction, and solving $\ii\hbar(d\hat\rho/dt)=\big[\hat{\mathcal{H}},\hat\rho\big]$ to the lowest order contribution, we find exactly the same expressions as above, but replacing $\big|\psi^j(\rb)\big|^2$ by the probability densities $\big\langle\rb|\hat\rho^j|\rb\big\rangle=(1/V)\sum_{\qb\qb'}\rho^j_{\qb\qb'}\ee^{\ii(\qb-\qb')\cdot\rb}$, based on which we can deal with electrons that have experienced decoherence before reaching the sample region.

An important point to consider is that bunched electrons are affected by Coulomb repulsion, which can increase the beam energy width and introduce undesired lateral deflections. For example, two 100\,keV electrons traversing a sample interaction region of length $L\sim10\,\mu$m with a relative longitudinal (transverse) separation distance of 1\,$\mu$m undergo a change in their energy (lateral deflection angle) of 14\,meV (0.1\,$\mu$rad). These values are still tolerable when probing visible and near-infrared optical excitations, but they increase linearly with $L$, becoming a limiting factor for propagation along the macroscopic beam column. We therefore anticipate that a strategy is needed to avoid them, such as introducing a large beam convergence angle ({\it i.e.}, large electron-electron distances except near the sampled region) or separating them by multiples of the optical period associated with the sampled excitation ({\it e.g.}, $4.1\,$fs for 1\,eV modes, corresponding to a longitudinal electron peak separation of 680\,nm at 100\,keV).

\begin{figure*}
\centering{\includegraphics[width=0.85\textwidth]{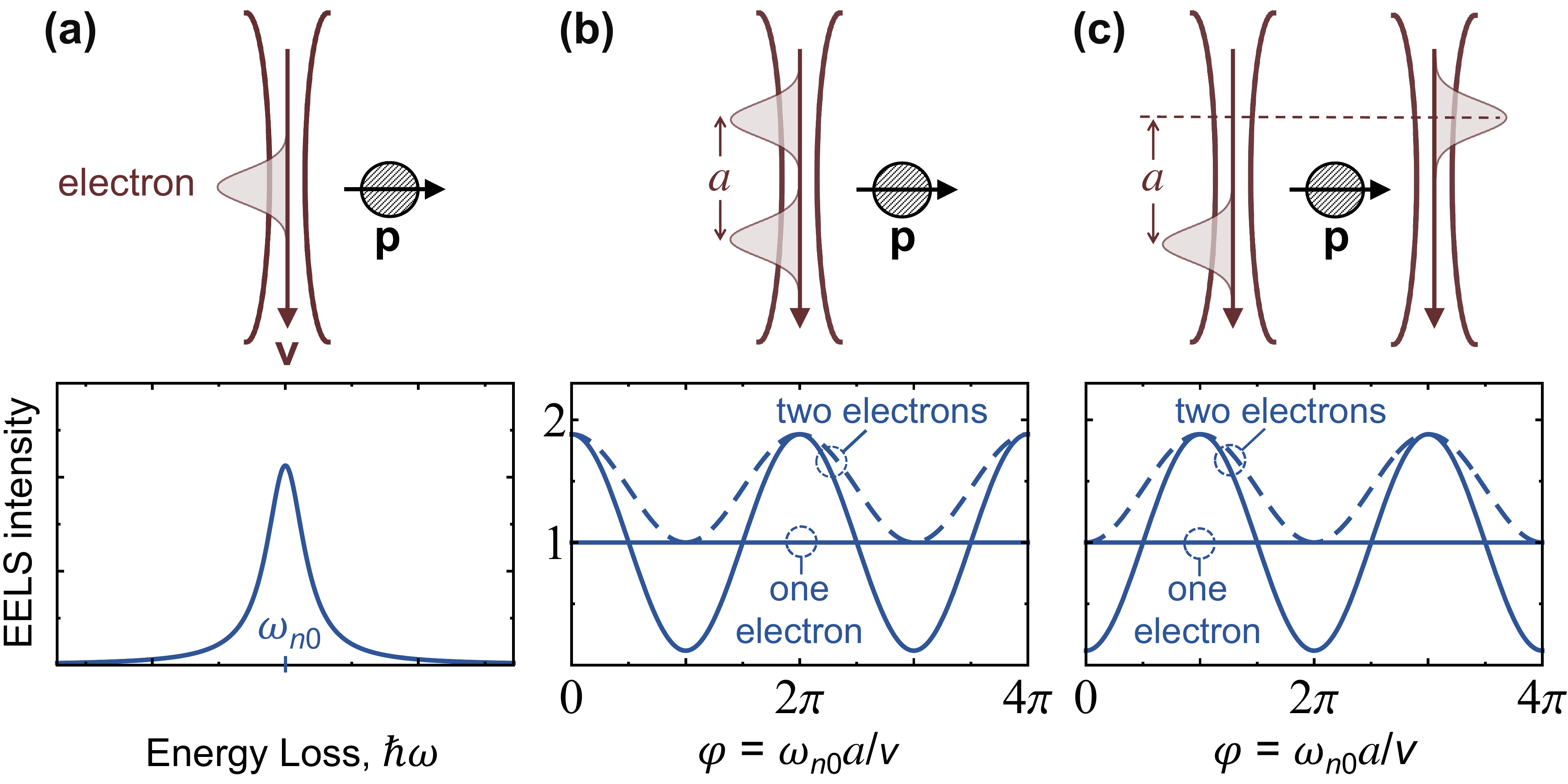}}
\caption{Interference in single- and double-electron interactions with a localized excitation. (a) Sketch of an electron wavepacket interacting with a nanoparticle (top) and typical EELS spectrum (bottom) dominated by one resonance of frequency $\omega_{n0}$ and polarization $\pb$ normal to the electron velocity $\vb$. (b) Interaction with two electron wavepackets separated by a longitudinal distance $a$. If the wavepackets are part of a single-electron wave function, the EELS probability is independent of $a$ (one-electron solid curve). With two electrons, each of them in a different wavepacket, the EELS intensity per electron oscillates with $\omega_{n0}a/v$ and presents a maximum at $a=0$ (two-electron solid curve). For two electrons with their wave functions equally shared among the two wavepackets, the oscillations with $a$ exhibit less profound minima (two-electron dashed curve). (c) Interaction with two electron wavepackets in symmetrically arranged beams. We find similar results as in (b), but now the two-electron probability displays a minimum at $a=0$. We consider wavepackets of width $\Delta$ defined by $\omega_{n0}\Delta/v=0.5$ (see Appendix). The EELS intensity is normalized to the result for uncorrelated electrons.}
\label{Fig5}
\end{figure*}

\subsection{Bunched and Dilute Electron-Beam Limits} We first consider $N$ electrons sharing the same form of the wave function, but separated by their arrival times $t_j=z_j/v$ at the region of interaction with the sample (also, see below an analysis of PINEM-modulated electrons, which belong to a different category), so we can write the incident wave functions as $\psi^j(\Rb,z)=\psi^0(\Rb,z-z_j)$, where $\psi^0$ is given by eq\ \ref{psi0}. Then, eq\ \ref{PNn} for the total excitation probability of mode $n$ reduces to
\begin{align}
&\Gamma_n^{\rm total}=N\Gamma_n^0+\left|Q_n^0\right|^2\sum_{j\neq j'}\ee^{\ii\omega_{n0}(z_{j'}-z_j)/v}
\label{bunch}
\end{align}
with $Q_n^0=\int d^3\rb\,\ee^{\ii\omega_{n0}z/v}\,|\psi^0(\rb)|^2\tilde{\beta}_n(\Rb)$ and $\Gamma_n^0$ given by eq\ \ref{P0n}. In addition, if the wave function displacements of all electrons satisfy $|z_j-z_{j'}|\ll v/\omega_{n0}$, neglecting linear terms in $N$, the sum in eq\ \ref{bunch} becomes $\approx N^2\left|Q_n^0\right|^2$, which can reach high values for large $N$, an effect known as superradiance when $n$ represents a radiative mode. We note that this effect does not require electrons confined within a small distance compared with the excitation length $v/\omega_{n0}$: superradiance is thus predicted to also take place for extended electron wave functions, provided all electrons share the same probability density, apart from some small longitudinal displacements compared with $v/\omega_{n0}$ (or also displacements by multiples of $v/\omega_{n0}$, see below); however, the magnitude of $Q_n^0$ will obviously decrease when each electron extends over several $v/\omega_{n0}$ spatial periods. Of course, if the electron density is further confined within a small region compared with $v/\omega_{n0}$ (or if it consists of a comb-like profile as given, for example, by eq\ \ref{psicomb}), we readily find $\Gamma_n^{\rm total}\approx N^2\Gamma_n^0$. Superradiance has been experimentally observed for bunched electrons over a wide range of frequencies \cite{SH1969,UGK98} and constitutes the basis for free-electron lasers \cite{AB04,EAA10,GIF19}.

In the opposite limit of randomly arriving electrons ({\it i.e.}, a dilute beam), with the displacements $z_j$ spanning a large spatial interval compared with $v/\omega_{n0}$ (even under perfect lateral alignment conditions), the sum in eq\ \ref{bunch} averages out, so we obtain $\Gamma_n^{\rm total}=N\Gamma_n^0$, and therefore, correlation effects are washed out.

\subsection{Superradiance with PINEM-Modulated Electrons} When $N$ electrons are modulated through PINEM interaction using the same laser (and neglecting $A^2$ corrections), their probability densities take the form
\[|\psi^j(\rb)|^2=|\psi_i^j(\rb)|^2\;|\mathcal{P}_d(\beta,\omega,z)|^2,\]
where the modulation factor $\mathcal{P}_d(\beta,\omega,z)$, defined in eq\ \ref{PPINEMd}, is shared among all of them and the PINEM coupling coefficient $\beta$ is taken to be independent of lateral position. Assuming well collimated e-beams, we consider the incident wave functions to be separated as $\psi_i^j(\rb)=\psi_\perp(\Rb)\psi_{i,\parallel}^j(z)$ ({\it i.e.}, sharing a common transverse component $\psi_\perp(\Rb)$ that is normalized as $\int d^2\Rb\,|\psi_\perp(\Rb)|^2=1$). Inserting these expressions into eqs\ \ref{PNn}-\ref{Mnj}, we find
\[\Gamma_n^{\rm total}=N\Gamma_n^0+|Q_n|^2\sum_{j\neq j'} M_n^jM_n^{j'*}\]
with
\[M_n^j=\int_{-\infty}^\infty dz\;\ee^{\ii\omega_{n0}z/v}\,|\psi_\parallel^j(z)\,\mathcal{P}_d(\beta,\omega,z)|^2,\]
where
\begin{align}
\Gamma_n^0&=\int d^2\Rb\;|\psi_{\perp}(\Rb)|^2 \,|\tilde{\beta}_n(\Rb)|^2, \nonumber\\
Q_n&=\int d^2\Rb\;|\psi_{\perp}(\Rb)|^2\tilde{\beta}_n(\Rb)
\nonumber
\end{align}
are transverse averages of the electron-sample coupling coefficient $\tilde{\beta}_n$. In general, the envelopes $|\psi_\parallel^j(z)|^2$ of the incident electrons are smooth functions that extend over many optical periods ({\it i.e.}, a large length $L$ compared with $v/\omega_{n0}$) and varies negligibly over each of them, so we can approximate
\begin{align}
M_n^j\approx M_n\equiv\lim_{L\to\infty}\frac{1}{L}\int_{-L/2}^{L/2} dz\;\ee^{\ii\omega_{n0}z/v}\,|\mathcal{P}_d(\beta,\omega,z)|^2. \nonumber
\end{align}
In this limit, $M_n$ is independent of the electron wave functions and arrival times, so it vanishes unless the sampled frequency $\omega_{n0}$ is a multiple of the PINEM laser frequency $\omega$. In particular, for $\omega_{n0}=m\omega$, where $m$ is an integer, using eq\ \ref{PPINEMd}, we find
\begin{align}
|M_n|&=\left|\sum_{l=-\infty}^\infty J_l(2|\beta|)\,J_{l+m}(2|\beta|)\,\ee^{4\pi\ii mld/z_T}\right| \nonumber\\
&=\big|J_m\big[4|\beta|\sin(2\pi md/z_T)\big], \label{MMnum}
\end{align}
where the second line is in agreement with ref \citenum{ZSF21} and directly follows from the first one by applying Graf’s addition theorem (eq\ (9.1.79) in ref\ \citenum{AS1972}). The total excitation probability then becomes
\begin{align}
\Gamma_n^{\rm total}=N\Gamma_n^0+N(N-1)\,|Q_n M_n|^2,
\label{N2pinem}
\end{align}
which contains an $N^2$ term ({\it i.e.}, superradiance). For tightly focused electrons, such that $|\psi_\perp(\Rb)|^2\approx\delta(\Rb-\Rb_0)$, we have $|Q_n|^2\approx\Gamma_n^0$, and consequently, eq\ \ref{N2pinem} reduces to $\Gamma_n^{\rm total}=\Gamma_n^0\;\left[N+N(N-1)\,|M_n|^2\right]$. This effect was predicted by Gover and Yariv \cite{GY20} by describing the electrons through their probability densities, treated as classical external charge distributions, and calculating the accumulated excitation effect, which is indeed independent of the arrival times of the electrons, provided they are contained within a small interval compared with the lifetime of the sampled mode $n$. Analogous cooperative multiple-electron effects were studied in the context of the Schwartz-Hora effect \cite{SH1969} by Favro {\it et al.}\cite{FFK1971}, who pointed out that a modulated "beam of electrons acts as a carrier of the frequency and phase information of the modulator and is able to probe the target with a resolution which is determined by the modulator". The obtained $N^2$ term thus provides a potential way of enhancing the excitation probability to probe modes with weak coupling to the electron. Incidentally, by numerially evaluating eq\ \ref{MMnum}, PINEM modulation using monochromatic light can be shown to yield \cite{paperarxiv6} $|M_n|^2\le34\%$, so additional work is needed in order to push this value closer to the maximum limit of $100\%$ obtained for $\delta$-function pulse trains.

\subsection{Interaction with Localized Excitations} For illustration purposes, we consider a laterally focused Gaussian electron wavepacket with probability density $|\psi^0(\rb)|^2\approx\delta(\Rb-\bb)\,\ee^{-z^2/\Delta^2}/(\sqrt{\pi}\Delta)$ interacting with a localized excitation of frequency $\omega_{n0}$ and transition dipole $\pb$ oriented as shown in Figure\ \ref{Fig5}a. The EELS probability is then described by a coupling coefficient that depends on $\pb$ and the direction of $\Rb$ as \cite{paper339} $\tilde{\beta}_0(\Rb)\propto\pb\cdot\hat{\Rb}$. Using these expressions for a single electron arranged in the two-wavepacket configurations of Figure\ \ref{Fig5}b,c, we find from eq\ \ref{P0n} an excitation probability $\Gamma_n^0=|\tilde{\beta}_n(\bb)|^2\propto|\pb|^2$ that is independent of the longitudinal ({\it i.e.}, along the beam direction) wavepacket separation $a$. In contrast, for two electrons with each of them in a different wavepacket, we find from eqs\ \ref{PNn}-\ref{Mnj}
\begin{align}
\Gamma_n^{\rm total}/2\Gamma_n^0=1\pm S\cos(\varphi),
\label{Pnlocal1}
\end{align}
where $\varphi=\omega_{n0}a/v$, $S=\ee^{-\omega_{n0}^2\Delta^2/2v^2}$, and the $+$ and $-$ signs apply to the configurations of Figures\ \ref{Fig5}b and \ref{Fig5}c, respectively (see Appendix). Interestingly, for two electrons with their wave functions equally shared among the two wavepackets, we also observe oscillations with $a$ as
\begin{align}
\frac{\Gamma_n^{\rm total}}{2\Gamma_n^0}=1+S\cos^2(\varphi/2)
\label{Pnlocal2}
\end{align}
in the $a\gg\Delta$ limit for the configuration of Figure\ \ref{Fig5}b (and the same expression with cos replaced by sin for Figure\ \ref{Fig5}c), which corresponds to the situation considered in eq\ \ref{bunch} for $z_j$ independent of $j$ and two electrons sharing the same wave function. In general, for $N$ laterally focused electrons ({\it i.e.}, a generalization of Figure\ \ref{Fig5}b), each of them having a wave function that is periodically distributed among $L$ wavepackets with separation $a$, we have
\begin{align}
\frac{\Gamma_n^{\rm total}}{N\Gamma_n^0}=1+\frac{N-1}{L^2}\;S\;\frac{\sin^2(L\varphi/2)}{\sin^2(\varphi/2)}
\label{Gtotn}
\end{align}
(see Appendix), which presents a maximum excitation probability $\Gamma_n^{\rm total}=N\,\left[1+(N-1)S\right]\,\Gamma_n^0$ (for $\varphi\rightarrow0$ or a multiple of $2\pi$) independent of the number of periods $L$.

\begin{figure}
\centering{\includegraphics[width=0.4\textwidth]{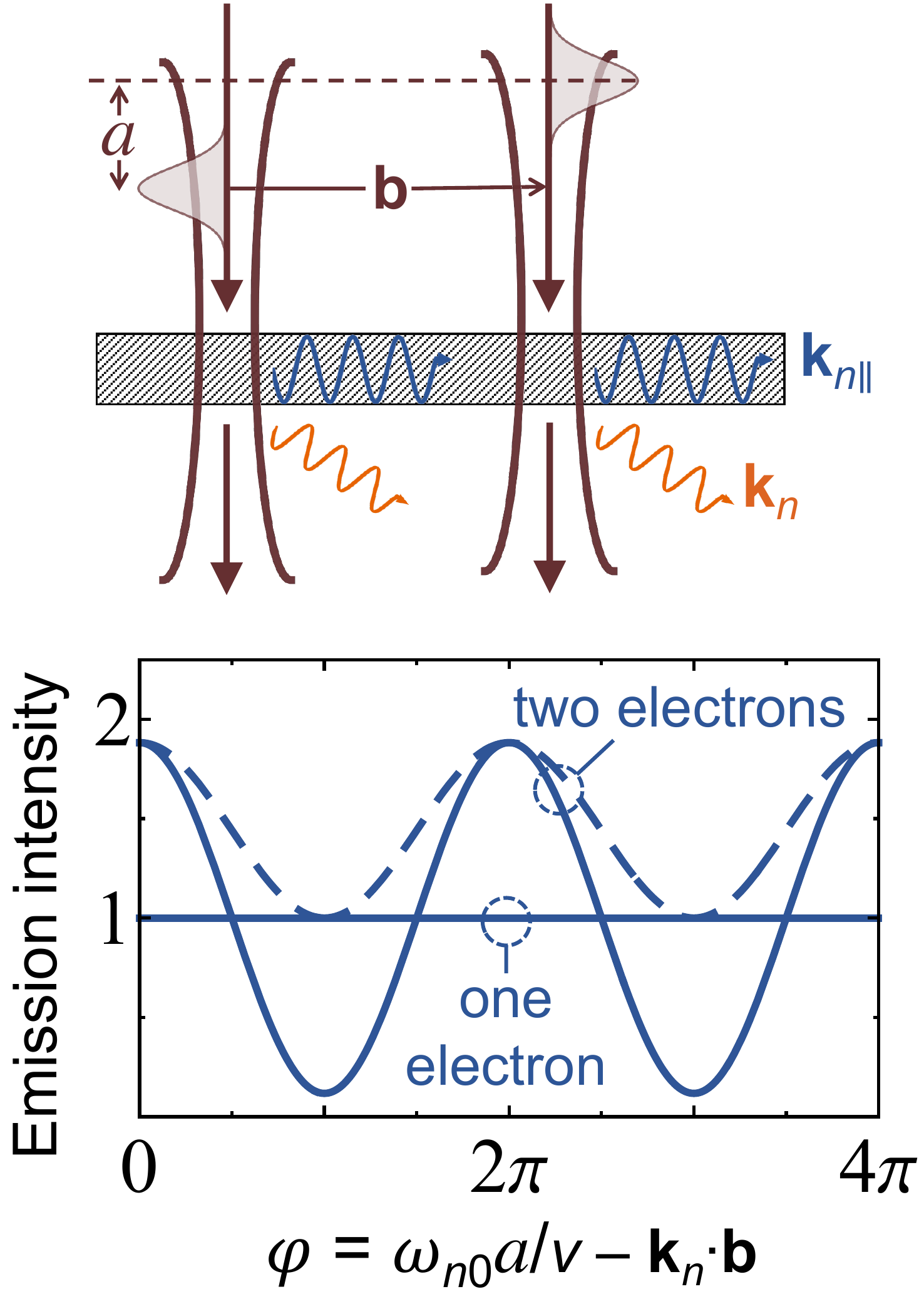}}
\caption{Interference in the interaction with delocalized modes. For the two-wavepacket beam configuration of Figure\ \ref{Fig5} and a sample that has lateral translational invariance, a single electron of split wave function emits in-plane polaritons and transition radiation with an intensity that is insensitive to the longitudinal and lateral wavepacket separations $a$ and $b$. This is in contrast to the emission intensity observed when each wavepacket is populated by one electron (two-electron solid curve) or when considering two electrons with each of them equally shared among the two wavepackets (two-electron dashed curve). We adopt the same beam parameters as in Figure\ \ref{Fig5} (see also Appendix).}
\label{Fig6}
\end{figure}

\subsection{Interference in the Emission of Photons and Polaritons} When the sample possesses lateral translational invariance, like in Figure\ \ref{Fig6}, the excited modes possess well-defined in-plane wave vectors $\kb_{n\parallel}$, so the coupling coefficients exhibit a simple spatial dependence, $\tilde{\beta}_n(\Rb)\propto\tilde{\beta}_n(0)\ee^{\ii\kb_{n\parallel}\cdot\Rb}$. Proceeding in a similar way as above for Gaussian wavepackets, we find no dependence on the wave function for single electrons, whereas for two electrons we obtain the same results as in eqs\ \ref{Pnlocal1} and \ref{Pnlocal2} with $\varphi$ redefined as $\omega_{n0}a/v-\kb_n\cdot\bb$. The emission probability thus oscillates with both longitudinal and lateral wavepacket displacements, $a$ and $\bb$, respectively, as illustrated in Figure\ \ref{Fig6}.

Incidentally, if the e-beam is laterally focused within a small region compared to $2\pi/k_{n\parallel}$, polaritons emitted to the left and to the right can interfere in the far field ({\it i.e.}, the final state $n$ is then comprising the detection system through which interference is measured by introducing an optical delay between the two directions of emission), while the interference is simply washed out as a result of lateral intensity averaging over the transverse beam profile if this extends over several polariton periods. This argument can be equivalently formulated in terms of the recoil produced on the electron due to lateral momentum transfer and the respective loss or preservation of {\it which way} information in those two scenarios, depending on whether such transfer is larger or smaller than the momentum spread of the incident electron \cite{KRA21}.

\subsection{Are Free Electrons Quantum or Classical Probes?} When examining a sample excitation of frequency $\omega_{n0}$ within a classical treatment of the electron as a point charge, the external source can be assimilated to a line charge with an $\ee^{\ii\omega_{n0}z/v}$ phase profile. The excitation strength by such a classical charge distribution coincides with $|\tilde{\beta}_n(\Rb)|^2$ (see eq\ \ref{betan}), where $\Rb$ gives the transverse position of the line. Actually, summing over all final states to calculate the EELS probability $\sum_n|\tilde{\beta}_n|^2\delta(\omega-\omega_{n0})$, we obtain a compact expression in terms of the electromagnetic Green tensor of the sample \cite{paper357} (eq\ \ref{EELSQM}, see detailed derivation in the Appendix), which is widely used in practical simulations \cite{paper149}. Extrapolating this classical picture to the configuration of Figure\ \ref{Fig6}, we consider two point electrons with lateral and longitudinal relative displacements, which directly yield an emission probability as described by eq\ \ref{Pnlocal1}. However, the classical picture breaks down for electrons whose wave functions are separated into several wavepackets: for single electrons, no classical interference between the emission from different wavepackets is observed, as the excitation probability reduces to a simple average of the line charge classical model over the transvese beam profile; likewise, for multiple electrons the excitation probability depends on the electron wave function in a way that cannot be directly anticipated from the classical picture ({\it cf.} solid and dashed curves in Figures\ \ref{Fig5} and \ref{Fig6}). The effect is also dramatic if the incident electrons are prepared in mutually entangled states, as discussed in a recent study \cite{KRA21_2}, while entangled electrons have also been proposed as a way to reduce beam damage in transmission electron microscopy \cite{OK14}.

The classical model provides an intuitive picture of interference in the CL emission from structured samples, such as in Smith-Purcell radiation \cite{SP1953} from periodic \cite{V1973b,HRS97}, quasiperiodic \cite{paper273}, and focusing \cite{RSR17} gratings. In our formalism, the coherent properties of the emitted radiation are captured by the $z$ integral in eq\ \ref{betan}, where the matrix element of the interaction Hamiltonian reduces to the electric field associated with the excited mode \cite{paper339}. In CL, the excited state $n$ refers to a click in a photon detector, and therefore, the sample must be understood as a complex system composed by the structure probed in the microscope, the optical setup, and the detector itself.

We remark that our results hold general applicability to any type of interaction Hamiltonian whose matrix elements $\langle n|\hat{\mathcal{H}}_1(\rb)|0\rangle$ are just a function of electron position $\rb$ (see eq\ \ref{betan}). This includes arbitrarily complex materials and their excitations, as well as the coupling to any external field. In particular, when describing the interaction with quantum electromagnetic fields through a linearized minimal-coupling Hamiltonian $\hat{\mathcal{H}}_1(\rb)\propto\hat\Ab(\rb)$, where $\hat\Ab(\rb)$ is the vector potential operator, the present formalism leads to the well-known EELS expression in eq\ \ref{EELSQM} (see derivation in the Appendix), which does account for coupling to radiation, and in particular, it can readily be used to explain the Smith-Purcell effect in nonabsorbing gratings \cite{paper149} ({\it i.e.}, when $\Gamma_{\rm CL}=\Gamma_{\rm EELS}$). This corroborates the generality of the present procedure based on treating the sample ({\it i.e.}, the universe excluding the e-beam) as a closed system, so its excitations are eigenstates of infinite lifetime. In a more traditional treatment of the sample as an open system, our results can directly be applied to excitations of long lifetime compared with the electron pulse durations. Additionally, coupling to continua of external modes can be incorporated through the Fano formalism \cite{F1961} to produce, for example, spectral CL emission profiles from the probabilities obtained for the excitation of confined electronic systems ({\it e.g.}, plasmonic nanoparticles).

We hope that this discussion provides some intuitive understanding on the role of the wave function in e-beam inelastic scattering, summarized in the statement that the excitation process by an individual swift electron (in EELS and CL) can be rigorously described by adopting the classical point-particle model, unless recoil becomes important ({\it e.g.}, for low-energy electrons or high-energy excitations). In contrast, the excitation by multiple electrons is affected by their quantum mechanical nature and depends on how their combined wave function is initially prepared. The predicted effects could be experimentally corroborated using few-electron pulses produced, for instance, by shaped laser pulses acting on photocathodes or {\it via} multiple ionization from ultracold atoms or molecules \cite{FFV17}. Besides its fundamental interest, the dependence of the excitation probability on the wave function for multiple electrons opens the possibility of realising electron-electron pump-probe imaging with an ultimate time resolution that is fundamentally limited by approximately half of the electron period $\pi/vq_0$ ({\it e.g.}, $\sim10^{-20}\,$s for 100\,keV electrons).

\begin{figure*}
\centering{\includegraphics[width=1.00\textwidth]{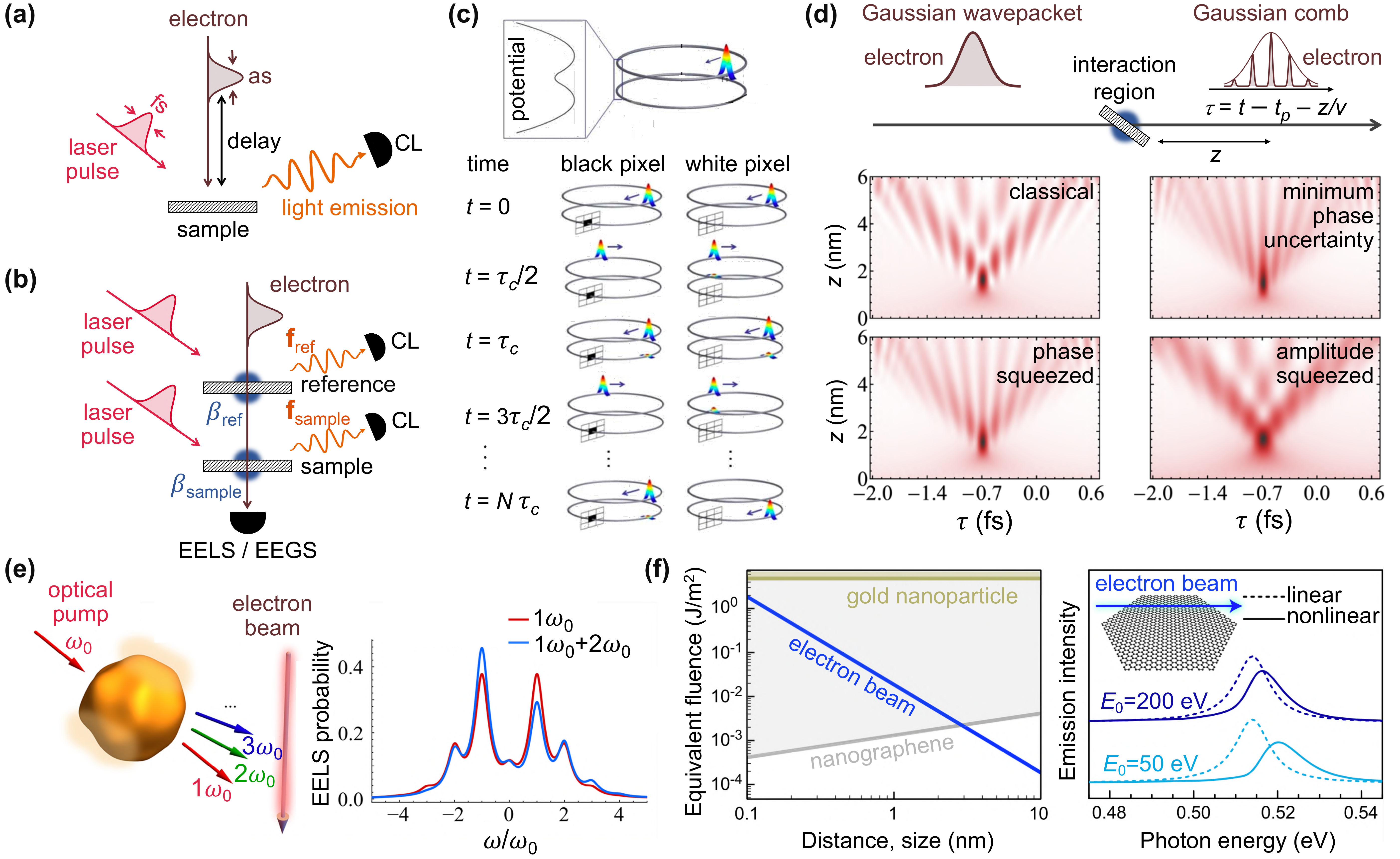}}
\caption{Future directions in photonics with electron beams.
(a) Combination of a fs laser pump synchronized with an attosecond electron pulse and detection of CL as an approach towards sub-{\AA}--attosecond--sub-meV resolution.
(b) Interferometric detection of a small sample object through EEGS measurements yielding the PINEM coupling coefficient $|\beta_{\rm ref}+\beta_{\rm sample}|^2\approx|\beta_{\rm ref}|^2+2{\rm Re}\{\beta_{\rm ref}^*\beta_{\rm sample}\}$, where the sample signal $\beta_{\rm sample}$ ($\ll1$) enters linearly and is amplified by an order-unity reference $\beta_{\rm ref}$. Alternatively, a similar scheme can be followed with the CL far-field intensity $I_{\rm CL}=|\fb_{\rm ref}+\fb_{\rm sample}|^2\approx|\fb_{\rm ref}|^2+2{\rm Re}\{\fb_{\rm ref}^*\cdot\fb_{\rm sample}\}$.
(c) Quantum electron microscopy for interaction-free imaging based on the quantum Zeno effect, whereby the presence of an object produces unity-order effects in the electron signal without the electron ever intersecting the sample materials. Adapted from ref\ \citenum{PY09}.
(d) Electron temporal compression after propagating a distance $z$ beyond the region of PINEM interaction (at time $t_p$) using classical and quantum light; the contour plots show the electron probability density as a function of propagation-distance-shifted time $\tau=t-t_p-z/v$ \cite{paper360}. Adapted from ref\ \citenum{paper360}.
(e) Sampling the nonlinear response of materials with nanoscale precision through the observation of harmonic-assisted asymmetry in the PINEM spectra. Adapted from ref\ \citenum{paper347}.
(f) Electron-beam-induced nonlinearities in small nanostructures, whereby low-energy electrons act equivalently to a high-fluence light pulse (left, for 25 eV electrons) and modify the EELS or CL spectra relative to the linear-interaction limit (right). Adapted from ref\ \citenum{paper350}.
}
\label{Fig7}
\end{figure*}

\section{Outlook and Perspectives}

We conclude this article with a succinct discussion of several promising directions for future research at the intersection of electron microscopy and photonics. The following is not an exhaustive list, but we hope that the reader can find in it some of the elements that are triggering a high degree of excitement in the nascent community gathered around this expanding field, including the promise for radical improvements in our way to visualize optical excitations with unprecedented space-time-energy resolution, as well as the opening of new directions in the study of fundamental phenomena.

\subsection{Towards Combined Sub-{\AA}--Attosecond--Sub-meV Resolution} PINEM-based UTEM is already in place to simultaneously combine nm--fs--sub-eV resolution inherited from focused e-beams, ultrafast optics, and EELS detection (see Figure\ \ref{Fig4} and references therein). The implementation of this technique in state-of-the-art aberration-corrected microscopes could push it further to the sub-{\AA} range, which, combined with fine tuning of the laser frequency, could lead to simultaneous sub-meV resolution via EEGS \cite{paper114,paper221}. Temporal resolution is then limited by the uncertainty principle $\sigma_E\sigma_t\ge\hbar/2\sim300\,{\rm meV}\times{\rm fs}$ relating the standard deviations of the electron pulse energy spread and time duration ($\sigma_E$ and $\sigma_t$, respectively) if the probe that is used to provide temporal resolution ({\it i.e.}, the compressed electron) is also energy-analyzed to resolve the excitation frequency through EELS. However, this limitation can be overcome if two different particles are employed to provide energy and time resolutions, respectively ({\it i.e.}, the uncertainty principle affects each of them individually, but not their crossed uncertainties). This possibility could be realized, for instance, by using single attosecond electron pulses to achieve time resolution with respect to a phased-locked optical pump, in combination with detection of the CL signal produced by the electron, as indicated by the red colored CL blob in Figure\ \ref{Fig1}a; sub-meV spectral resolution could then be gained through optical spectroscopy (see Figure\ \ref{Fig7}a). Besides the technical challenge of combining fs-laser and attosecond-electron pulses \cite{MB20}, detection of CL emission can be difficult because it may be masked by light scattered from the laser, so it needs to be contrasted with the optical signal observed in separate measurements using only electrons or laser irradiation, or alternatively, laser scattering could be interferometrically removed at the light spectrometer.

\subsection{Non-Invasive Imaging: Interferometric and Quantum Electron Microscopies} Sample damage is a major source of concern in electron microscopy, particularly when investigating soft and biological materials. Besides cooling the sample to make it more resistant (cryogenic electron microscopy \cite{F02_2}), various strategies can be followed to combat this problem, essentially consisting in enhancing the signal contrast produced by the specimen with a minimum interaction with the electrons. This is the principle underlying the proposed quantum electron microscope \cite{PY09} (see Figure\ \ref{Fig7}c), inspired in a previously explored form of interaction-free optical microscopy \cite{KWM99}, and consisting in initially placing the electron in a cyclic free path (upper potential well) that has a small probability amplitude $T$ of transferring into a second cyclic path (lower potential well) during a cycle time period $\tau_c$. The second path is taken to intersect the sample, and therefore, the quantum Zeno effect resolves the question whether a given pixel contains material or is instead empty: when the lower path passes through a {\it filled} sample pixel, the electron wave function collapses, so the overall transfer into this path after a time $N\tau_c$ ({\it i.e.}, after $N$ roundtrips) reduces to $\sim N|T|^2$; in contrast, when the lower path passes through an {\it empty} sample pixel, the accumulated transfer of probability amplitude becomes $\sim NT$, and the transferred probability is instead $\sim|NT|^2$. Consequently, for large $N$ and small $|T|$, such that $|NT|^2\sim1$, detection of the electron in the upper path indicates that a filled pixel is being sampled, involving just a marginal probability $\sim N|T|^2$ of electron-sample collision; on the contrary, an empty sample pixel is revealed by a depletion $\sim|NT|^2$ in the electron probability associated with the upper path, equally avoiding sample damage because there is no material to collide. An international consortium is currently undertaking the practical implementation of this challenging and appealing form of microscopy \cite{KHK16}. An extension of this idea to incorporate the detection of sample optical excitations and their spectral shapes would be also desirable in order to retrieve valuable information for photonics.

Interferometry in the CL signal offers a practical approach to study the response of small scatterers by using the electron as a localized light source that is positioned with nanometer precision in the neighborhood of the object under study \cite{paper341,SAG20}. In a related development, CL light produced by an engineered metamaterial reference structure has been postulated as a source of ultrafast focused light pulses that could be eventually combined with the exciting electron in a pump-probe configuration \cite{T18,TMG19}. These studies inspire an alternative way of reducing sample damage (Figure\ \ref{Fig7}b, CL emission), also in analogy to infrared SNOM \cite{HTK02}: by making the electron to traverse a reference structure ({\it e.g.}, a thin film), followed by interaction with the sample, the CL far-field amplitudes $\fb_{\rm ref}$ and $\fb_{\rm sample}$ produced by these events are coherently superimposed ({\it i.e.}, both of them maintain phase coherence, just like the emission emanating from the different grooves of a grating in the Smith-Purcell effect \cite{SP1953}), giving rise to a CL intensity $I_{\rm CL}=|\fb_{\rm ref}+\fb_{\rm sample}|^2\approx|\fb_{\rm ref}|^2+2{\rm Re}\{\fb_{\rm ref}^*\cdot\fb_{\rm sample}\}$, where the sample signal in the second term is amplified by a stronger reference signal ({\it i.e.}, we take $|\fb_{\rm ref}|\gg|\fb_{\rm sample}|$) that can be calibrated {\it a priori}. This strategy can provide a large sample signal compared with direct (unreferenced) CL detection ({\it i.e.}, $|2{\rm Re}\{\fb_{\rm ref}^*\cdot\fb_{\rm sample}\}|\gg|\fb_{\rm sample}|^2$), and thus, the electron dose needed to collect a given amount of information is reduced, or alternatively, there is some flexibility to aim the e-beam a bit farther apart from the specimen to reduce damage.

In the context of UTEM, the demonstration of coherent double-PINEM interactions \cite{EFS16} opens a similar interferometric avenue to reduce sample damage by associating them with reference and sample structures (Figure\ \ref{Fig7}b). The PINEM spectrum responds to the overall coupling strength $|\beta_{\rm ref}+\beta_{\rm sample}|^2$ (see the discussion on the addition property of $\mathcal{P}_0(\beta,\omega,z)$ after eq\ \ref{PPINEM}), which contains an interference term $2{\rm Re}\{\beta_{\rm ref}^*\beta_{\rm sample}\}$ that can again amplify a weak PINEM signal from an illuminated sample by mixing it with a strong reference. This effect has also been studied in connection with the interaction between a free electron and a two-level atom \cite{PG19,ZSF21,RGM21}, where the inelastic electron signal is found to contain a component that scales linearly with the electron-atom coupling coefficient if the electron wave function is modulated and the atom is prepared in a coherent superposition of ground and excited states that is phase-locked with respect to the electron modulation (in contrast to a quadratic dependence on the coupling coefficient if the atom is prepared in the ground state). We remark the necessity of precise timing ({\it i.e.}, small uncertainty compared with the optical period of the excitation) between the electron modulation and the amplitudes of ground and excited states in the two-level system. This condition could be met in the double-PINEM configuration, giving rise to an increase in sensing capabilities, so that a smaller number of beam electrons would be needed to characterize a given object ({\it e.g.}, a fragile biomolecule).

It should be noted that, despite their appeal from a conceptual viewpoint, individual two-level Fermionic systems present a practical challenge because the transition strength of these types of systems is typically small ({\it e.g.}, they generally contribute with $\lesssim1$ electrons to the transition strength, as quantified through the f-sum rule \cite{PN1966,paper267}), and in addition, coupling to free electrons cannot be amplified through PINEM interaction beyond the level of one excitation, in contrast to bosonic systems ({\it e.g.}, linearly responding plasmonic and photonic cavities, which can be multiply populated). Nevertheless, there is strong interest in pushing e-beam spectroscopies to the single-molecule level, as recently realized by using high-resolution EELS for mid-infrared atomic vibrations \cite{HHP19,HKR19,HRK20,ZZW21} (see Figure\ \ref{Fig2}d), which are bosonic in nature and give rise to measurable spectral features facilitated by the increase in excitation strength with decreasing frequency. However, e-beam-based measurement of valence electronic excitations in individual molecules, which generally belong to the two-level category, remains unattained with atomic-scale spatial resolution. In this respect, enhancement of the molecular signal by coupling to a nanoparticle plasmon has been proposed to detect the resulting hybrid optical modes with the e-beam positioned at a large distance from the molecule to avoid damage \cite{KNA18}. The interferometric double-PINEM approach could provide another practical route to addressing this challenge. The $N^2$ excitation predicted for PINEM-modulated electrons \cite{GY20} (see eq\ \ref{N2pinem}) is also promising as a way to amplify specific probed excitation energies while still maintaining a low level of damage $\propto N$.

Interferometric CL and PINEM approaches should enable the determination of the phase associated with the emitted and induced optical near fields, respectively. In CL, this could be achieved without modifying the e-beam components of the microscope by introducing a tunable optical delay line in the light component emanating from the reference structure before mixing it with the sample component. In PINEM, the delay line could be incorporated in the laser field  illuminating the reference structure. The quantities to be determined are the complex scattering amplitude (CL) and the near field (PINEM), which are actually two sides of the same coin, related through eq\ \ref{betaCL2}. If the reference signal is well characterized and in good correspondence with theory ({\it e.g.}, transition radiation from a thin film \cite{YAT01}), this procedure should enable the determination of the frequency-dependent optical phase. In addition, self-interference of the CL signal ({\it e.g.}, by mixing different emission directions through a bi-prism) could provide a simple method to measure the angular dependence of the far-field complex amplitude, while the interferometric detection discussed above can supply the missing information from the spectral dependence of the phase.

\subsection{Interference between E-Beam and External Light Excitations} Recent reports \cite{paperarxiv3,paperarxiv6} have revealed that CL emission can interfere with external light that is synchronized with the electron wave function. This effect has been found to be controlled by the same coherence factors $M_n^j$ that intervene in the interference among different beamed electrons (eq\ \ref{Mnj}). An extension of those results to general excitations in the specimen can be obtained by following the procedure used in the derivation of eq\ \ref{P0n}, but including the interaction with a weak ({\it i.e.}, acting linearly) classical field ({\it e.g.}, laser light) of finite temporal duration. The latter can be introduced through an additional time-dependent interaction Hamiltonian
\begin{align}
\hat{\mathcal{H}}_2(t)=\int \frac{{\rm d}\omega}{2\pi}\,\hat{\mathcal{H}}_2(\omega)\ee^{-\ii\omega t}
\nonumber
\end{align}
This expression automatically implies synchronization of the classical field and the beam electrons by selecting a common time origin. Expanding the wave function of the system as in eq\ \ref{defpsi}, we find the post-interaction coefficients given by eq\ \ref{alphainfinity}, but now supplemented by an additional term $(-\ii/\hbar)\big\langle n\big|\hat{\mathcal{H}}_2(\omega_{n0})\big|0\big\rangle\alpha_\qb^0$. From here, proceeding in a way analogous to the derivation of eq\ \ref{P0n} in the Appendix, the excitation probability of a mode $n$ is found to be
\begin{align}
&\Gamma_n^0=
\int d^3\rb\;\big|\psi^0(\rb)\big|^2 \,\bigg|\tilde{\beta}_n(\Rb)\ee^{\ii\omega_{n0}z/v}+\beta_n^{\rm field}\bigg|^2 \nonumber\\
 \nonumber\\
&=\int d^3\rb\;\big|\psi^0(\rb)\big|^2 \,\big|\tilde{\beta}_n(\Rb)\big|^2+\big|\beta_n^{\rm field}\big|^2 +2\Ree\bigg\{\beta_n^{{\rm field}*}\,Q_n^0\bigg\}
\nonumber
\end{align}
where
\begin{align}
\beta_n^{{\rm field}*}=\frac{1}{\hbar}\big\langle n\big|\hat{\mathcal{H}}_2(\omega_{n0})\big|0\big\rangle
\nonumber
\end{align}
is an excitation amplitude associated with the external classical field, whereas $Q_n^0$ is defined in eq\ \ref{Qj}. Finally, following the same approach as in the derivation of eqs\ \ref{PNn}-\ref{Mnj} in the Appendix, we find an extension of this result to e-beams consisting of multiple distinguishable electrons:
\begin{align}
\Gamma_n^{\rm total}
=&\sum_j\int d^3\rb\;\big|\psi^j(\rb)\big|^2 \,\big|\tilde{\beta}_n(\Rb)\big|^2+\sum_{j\neq j'}Q_n^jQ_n^{j'*} \nonumber\\
&+\big|\beta_n^{\rm field}\big|^2 +2\sum_j\Ree\bigg\{\beta_n^{{\rm field}*}\,Q_n^j\bigg\}
\label{EEQ1}
\end{align}
where $j$ and $j'$ are electron labels. We thus confirm that the synchronized interactions between different electrons and light with a sample are both governed by the coherence factors defined in eqs\ \ref{Qj} and \ref{Mnj}. When the excitation mode corresponds to an emitted photon, this equation produces the angle- and frequency-dependent far-field photon probability
\begin{widetext}
\begin{align}
&\frac{d\Gamma_{ \rm rad}}{d\Omega_{\rr_\infty}d\omega}=\frac{c}{4\pi^2\hbar\omega}\Bigg{\{} \sum_j\int d^2\Rb\, M_0^j(\Rb) |\fb_{\rr_\infty}^{\rm CL}(\Rb,\omega)|^2 \nonumber\\
&\quad+|\fb_{\rr_\infty}^{\rm scat}(\omega)|^2+2\sum_j\int d^2\Rb\, {\rm Re}\left\{M_{\omega/v}^j(\Rb)\; \fb_{\rr_\infty}^{{\rm CL}*}(\Rb,\omega)\cdot \fb_{\rr_\infty}^{\rm scat}(\omega)\right\} \nonumber\\
&+\sum_{j\neq j'}
\left[\int d^2\Rb\, M_{\omega/v}^j(\Rb)\fb_{\rr_\infty}^{{\rm CL}*}(\Rb,\omega)\right]
\left[\int d^2\Rb'\, M_{\omega/v}^{j'*}(\Rb')\fb_{\rr_\infty}^{\rm CL}(\Rb',\omega)\right]
\Bigg{\}}
\label{EEQ2}
\end{align}
\end{widetext}
which is derived in ref\ \cite{paperarxiv6} from an alternative quantum electrodynamics formalism and constitutes an extension of eq\ \ref{anothereq} to include the simultaneous interaction with multiple electrons and an external light field. Here, the excitation frequency is denoted $\omega=\omega_{n0}$, the coherence factors are renamed as $M_{\omega/v}^j(\Rb)\equiv M_n^j(\Rb)$ (see eq\ \ref{Mnj}), and the far-field amplitude component $\fb_{\rr_\infty}^{\rm scat}(\omega)$ refers to the scattered laser field arriving at the same photon detector as the CL emission, either after scattering at the sample or directly from the employed laser. We obtain eq\ \ref{EEQ2} from eq\ \ref{EEQ1} by multiplying by $\delta(\omega-\omega_{n0})$, making the transformations $\tilde\beta_n(\Rb)\rightarrow\sqrt{c/4\pi^2\hbar\omega}\;\fb_{\rr_\infty}^{{\rm CL}*}(\Rb,\omega)$ and $\beta_n^{\rm field}\rightarrow\sqrt{c/4\pi^2\hbar\omega}\;\fb_{\rr_\infty}^{{\rm scat}*}(\omega)$, and summing over modes $n$ that contribute to the emission direction $\rr_\infty$. Obviously, in order to observe the interference between CL and laser light, the latter has to be dimmed, so that both of them have commensurate amplitudes, as extensively discussed in ref\ \cite{paperarxiv6}. The coherence factor $M_{\omega/v}^j(\Rb)$ determines the ability of each electron $j$ to interfere with synchronized light. This factor is maximized ($\big|M_{\omega/v}^j(\Rb)\big|\rightarrow1$) in the point-particle limit (see discuss above). This analysis reveals that temporally compressed electrons act as partially coherent, localized sources of excitation ({\it e.g.}, CL emission), tantamount to the external light, but with the faculty of acting with sub-nm spatial precision. Besides the prospects opened by these findings to control nanoscale optical excitations, this approach offers an alternative way of determining the absolute magnitude and phase of $\fb_{\rr_\infty}^{\rm CL}$ through the interference term in the above equation.

Incidentally, we remark again that the above expressions are directly applicable to electrons prepared in mixed states by substituting $\big|\psi^j(\rb)\big|^2$ by the electron probability density (see above).

\subsection{Manipulation of the Quantum Density Matrix Associated with Sample Modes} In addition to the aforementioned implementations of shaped electron beams for microscopy and imaging, the modulated electron wave function has been investigated as a means to manipulate the quantum state of confined optical excitations. This is relevant because of its potential to create states of light with nontrivial statistics, enabling exciting applications in quantum computing \cite{KMN07}, metrology \cite{GLM11}, and information \cite{WFP17}. An initially separable joint electron-sample state is generally brought to a complex entangled state after interaction, which upon partial tracing and projection over the electron degrees of freedom, allows us to modify the sample density matrix. Obviously, a wider range of sample states could be accessed by controlling the incoming electron density matrix, for example, through PINEM interaction with nonclassical light \cite{paper360} (see below). For a general initial electron-photon (e-p) density matrix $\rho_{\rm e,p}^i$, the joint final state after interaction can be written as $\rho_{\rm e,p}^f=\hat{\mathcal{S}}\rho_{\rm e,p}^i\hat{\mathcal{S}}^\dagger$ in terms of the scattering operator $\hat{\mathcal{S}}$. If the electron is not measured, the resulting photonic density matrix is obtained through the partial trace over electron degrees of freedom, $\rho^{\rm no-meas}_{\rm p}={\rm Tr}_{\rm e}\{\rho^f_{\rm e,p}\}$. When the sample is initially prepared in its ground state, the diagonal elements of $\rho^{\rm no-meas}_{\rm p}$ define a Poissonian distribution, regardless of the incident electron wave function \cite{paper360}, while off-diagonal terms exhibit a pronounced dependence that can potentially be measured through optical interferometry \cite{paperarxiv3} and direct mixing of CL and laser light scattering \cite{paperarxiv6}. Incidentally, in the point-particle limit for the electron, the interaction is equivalent to excitation of the sample by a classical current, which is known to transform an initial coherent state ({\it e.g.}, the sample ground state) into another classical coherent state \cite{GL91} (the excited sample). In contrast, if the electron is measured ({\it i.e.}, only instances of the experiment with a given final electron state $|\qb\rangle$ are selected), the interaction-induced e-p entanglement leads to a wide set of optical density matrices $\rho^{\rm meas}_{\rm p}={\rm Tr}_{\rm e}\{|\qb\rangle\langle\qb|\rho^f_{\rm e,p}\}\neq \rho^{\rm no-meas}_{\rm p}$ that can be post-selected through the detection of a transmitted electron with, for example, a specific wave vector $\qb$; obviously, using more than one electron further increases the range of possible outcomes. Single-photon generation triggered by energy-momentum-resolved transfers from an electron to a waveguide constitutes a trivial example of this strategy \cite{paper180}. This approach has also been proposed to produce thermal, displaced Fock, displaced squeezed, and coherent sample states \cite{HRN21}.

\subsection{Manipulation of the Electron Density Matrix} If no measurement is performed on the sample, interaction with the electron modifies the density matrix of the latter, which becomes $\rho^f_{\rm e}={\rm Tr}_{\rm p}\left\{\rho^f_{\rm e,p}\right\}$. For example, after PINEM interaction with laser light, we find (going to the Schr\"odinger picture) $\rho^f_{\rm e}(\rb,\rb',t)=\psi(\rb,t)\psi^*(\rb',t)$, where the wave function $\psi(\rb,t)$ (eq\ \ref{psiPINEM}) is controlled by a single coupling parameter $\beta$ (eq\ \ref{beta}). Also, the tranformation of a general incident density matrix $\rho^i(\rb,\rb',t)$ is mediated by the factors defined in eq\ 15 as
\begin{align}
&\rho^f(\rb,\rb',t) \nonumber\\
&=\mathcal{P}_d[\beta(\Rb),\omega,z-vt]\;\mathcal{P}^*_d[\beta(\Rb'),\omega,z'-vt]\;\rho^i(\rb,\rb',t).
\nonumber
\end{align}
More complex forms of $\rho^f_{\rm e}$ are obtained when using nonclassical light. In this respect, recent advances in quantum light sources ({\it e.g.}, squeezed light generation \cite{AGM16}) provide a practical way to induce nonclassical sample states, which in turn modulate the electron density matrix through PINEM-like interaction \cite{paper360}. We illustrate this idea by showing in Figure\ \ref{Fig7}d the diagonal part of the density matrix ({\it i.e.}, the electron probability density) for both laser and nonclassical illumination. When the phase uncertainty in the light state is decreased (phase-squeezed and minimum-phase-uncertainty \cite{KK93} optical states), the electron density peaks are found to be more compressed in time, and in addition, because of conservation of the total probability, a complementary elongation takes place along the propagation direction. In contrast, the opposite trend is observed when using amplitude-squeezed light. In the limit of illumination with maximum phase uncertainty, such as Fock and thermal optical states, the electron does not undergo compression because there is no coherence among sample states of different energy \cite{paper360}.

If the length of the e-beam--specimen interaction region is sufficiently small as to assume that eq\ \ref{nonrecoil} holds during the passage of the electron, the real-space representations of the initial and final electron density matrices (before and after interaction) depend on time as $\rho^{i,f}(\rb-\vb t,\rb'-\vb t)$. Then, after linear interaction with a specimen prepared in the ground state, these quantities are related as
\begin{align}
&\rho^f(\rb-\vb t,\rb'-\vb t) \nonumber\\
&=\exp\big[K(\Rb,\Rb',z-z')\big]\;\rho^i(\rb-\vb t,\rb'-\vb t)
\nonumber
\end{align}
where
\begin{align}
&K(\Rb,\Rb',z-z')=\frac{2e^2}{\hbar} \nonumber\\
&\times\int_0^\infty d\omega\int_{-\infty}^\infty {\rm d}z''\int_{-\infty}^\infty {\rm d}z'''\;\ee^{\ii\omega(z''-z''')/v}\nonumber\\
&\times\bigg[\ee^{-\ii\omega(z-z')/v}
\;2\Imm\big\{-G_{zz}(\Rb,z'',\Rb',z''',\omega)\big\} \nonumber\\
&\quad-\ii G_{zz}(\Rb,z'',\Rb,z''',\omega)+\ii G_{zz}^*(\Rb',z'',\Rb',z''',\omega)\bigg]
\nonumber
\end{align}
for $\vb$ along $z$. We have derived a linearized form of this expression ({\it i.e.}, with $\ee^K$ substituted by $1+K$) only assuming time reversal symmetry and the nonrecoil approximation as a direct extension of the techniques used in the Appendix when proving eqs\ \ref{EELSQM} and \ref{Gammafi}. The full result (with $\ee^K$) was obtained elsewhere within a quantum-electrodynamics formalism \cite{paper357}. Reassuringly, we have $K(\Rb,\Rb,0)=0$, so the norm $\int d^3\rb\,\rho(\rb,\rb)=1$ is preserved. In addition, the property $K^*(\Rb,\Rb',z-z')=K(\Rb',\Rb,z'-z)$ guarantees the Hermiticity of the transformed density matrix. We note that the $\Imm\{\dots\}$ term originates in inelastic scattering, while the remaining two terms are associated with elastic processes from the electron viewpoint, which are essential to conserve the norm.

For completeness, we note that, incorporating in eq\ \ref{nonrecoil} the lowest-order nonrecoil correction (i.e., $\varepsilon_\qb-\varepsilon_{\qb'}\approx\vb\cdot(\qb-\qb')+(\hbar/2\me\gamma^3)\big(|\qb-\qb_0|^2-|\qb'-\qb_0|^2\big)$ with $\qb_0=\me\vb\gamma/\hbar$), free electron propagation over a distance $d$ transforms the density matrix as
\begin{align}
&\rho^f(\Rb,z,\Rb',z',t) \nonumber\\
&=\int_{-\infty}^\infty {\rm d}z''\int_{-\infty}^\infty {\rm d}z'''\;T(z-z'',z'-z''')\;\rho^i(\Rb,z'',\Rb',z''',t)
\nonumber
\end{align}
with $T(z,z')=(-\ii\gamma^2q_0/2\pi d)\,\exp\big[(\ii\gamma^2q_0/2d)(z^2+z'^2)\big]$. In particular, this procedure readily yields eq\ \ref{PPINEM} from eq\ \ref{PPINEMd}.

\subsection{Nanoscale Sampling of the Nonlinear Optical Response} Electron beams potentially grant us access into the nonlinear response of materials with unprecedented nanoscale spatial resolution. Specifically, PINEM offers a possible platform to perform nonlinear nanoscale spectroscopy \cite{paper347} (Figure\ \ref{Fig7}e): under intense laser pulse irradiation, the sample can generate evanescent near fields not only at the fundamental frequency but also at its harmonics, which produce a departure from the gain-loss symmetry in the resulting EELS spectra. These types of asymmetries have already been demonstrated by performing PINEM with simultaneous $\omega$ and $2\omega$ external irradiation \cite{PRY17} ({\it i.e.}, through a combination of two PINEM interactions at such frequencies, as described by eq\ \ref{psiPINEM}, but with the $2\omega$ component now produced by external illumination having phase coherence relative to the $\omega$ laser field).

At lower kinetic energies, electrons produce an increasingly stronger perturbation on the sample, which has been speculated to eventually trigger a measurable nonlinear material response \cite{paper350}. The idea is that the electron acts as a relatively high-fluence optical pulse (Figure\ \ref{Fig7}f, left), so the resulting nonlinear field emanating from the sample could be traced through the shift in spectral features revealed by EELS or CL as the e-beam velocity or impact parameter are scanned (Figure\ \ref{Fig7}f, right).

In a related context, nanoscale ultrafast probing could eventually assist the exploration of quantum nonlinearites, such as those imprinted on bosonic cavity modes due to hybridization with two-level systems ({\it e.g.}, quantum emitters), which have been a recurrent subject of attention in recent years \cite{BHA05,DSF10,HCS15,paper176,paper339}.

\subsection{Optical Approach to Electron-Beam Aberration Correction} Advances in electron microscopy have been fuelled by a sustained reduction in e-beam aberrations and energy spread. In particular, both aberration-correction and lateral beam shaping rely on our ability to control the lateral electron wave function. This can be done with great precision using static microperforated plates, which, for example, enable the synthesis of highly chiral vortex electron beams \cite{VTS10,MAA11}. Dynamical control is however desirable for applications such as fast tracking of sample dynamics. Substantial progress in this direction is being made through the use of perforated plates with programable potentials that add a position-dependent electric Aharonov-Bohm phase to the electron wave function \cite{VBM18}. In a separate development, intense laser fields have been used to optically imprint a ponderomotive phase on the electrons \cite{MJD10,SAC19,ACS20} ({\it i.e.}, as described by eq\ \ref{phase}). Combined with UTEM and structured illumination, one could use strong, spatially modulated lasers to imprint an on-demand transverse phase profile on the electron wave function in order to correct aberrations and customize the focal spot profile. This general approach has been theoretically explored through PINEM interaction for light reflected on a continuous thin foil \cite{paper351}, as well as by relying on the free-space ponderomotive elastic phase \cite{paperarxiv4}. A recent study also proposes the use of PINEM interactions with spectrally shaped light pulses to reduce e-beam energy spreading \cite{RK20}. These advances constitute promising directions to enhance our control over the wave function of free electrons for application in improved e-beam-based, spectrally resolved microscopies.

\subsection{Nanoscale Electron-beam Photon Sources} By interacting with material boundaries, the evanescent field carried by electrons is transformed into propagating CL light emission. This effect has been extensively exploited to produce efficient light sources \cite{LCS16,CTM20}, for example with the e-beam flying parallel to a grating surface (Smith-Purcell effect \cite{SP1953,paper027,paper252,paper273,RKT19}), where superradiance ({\it i.e.}, when the emission intensity scales quadratically with the e-beam current) has been demonstrated in the generation of THz radiation \cite{UGK98}. Electron wiggling caused by periodic structures is equally used in undulators at synchrotrons, while a nanoscale version of this effect has also been proposed \cite{WKI16}. A particularly challenging task is the production of X-ray photons with nanometer control, which recent studies have tackled following different strategies, such as through the simultaneous generation of polaritons in a nonlinear two-quanta emission process \cite{RWJ19}, or by an atomic-scale version of the Smith-Purcell effect using atomic planes in van der Waals materials as the periodic structure \cite{paper356}. Additionally, a quantum klystron has recently been proposed based on spatially modulated intense electron beams in a PINEM-related configuration followed by free-space propagation, giving rise to a periodic train of electron bunches that could trigger superradiance from two-level emitters \cite{RHS21}, in analogy to the intriguing Schwartz-Hora effect \cite{SH1969,FFK1971}, which modern technology could perhaps revisit.

\subsection{Towards Free-Space Nanoelectronics at Low Kinetic Energies} In nanophotonics, there is a plethora of photon sources that can be integrated in nanostructured environments to control the flow of light for information processing, sensing, and other applications. When using free electrons instead of photons, things become more complicated because of the unavailability of nanoscale sources. As a preliminary step to fill this gap, multiphoton photoemission amplified by strong plasmonic field enhancement at the nm-sized tips of metallic nanoparticles has been demonstrated to provide a localized source of free electrons that can be generated using relatively weak light intensities down to the continuous-wave limit \cite{paper310}. Free-space nanoelectronics, consisting in moulding the flow of these electrons through nanostructured electric-potential and magnetic-field landscapes, thus emerges as an appealing research frontier with applications in micron-scale free-electron spectroscopy for sensing and detection devices. 

In a parallel approach, electrical and magnetic manipulation of ballistic electrons has recently been achieved in graphene \cite{CHE16,LGR17,BHS17,BCS17} and other 2D materials \cite{BSB19}, sharing some of the properties of free electrons, including the possibility of generating single-electron wavepackets \cite{FSH13}. Based on these developments, we envision the implementation of photon-free spectroscopy performed within 2D material devices, whereby electrical generation and detection of inelastically scattered ballistic electrons provides spectral information on the surrounding environment. A recent exploration of this idea has resulted in the proposal of ultrasensitive chemical identification based on electrical detection of EELS-like vibrational fingerprints from analytes placed in the vicinity of a 2D semiconductor exposed to a nanostructured potential landscape that could be achieved using existing gating technology \cite{paper349}.

\section{Appendix} 
\renewcommand{\thesection}{A} 
\renewcommand{\theequation}{A\arabic{equation}} 


\subsection{Expressing the EELS Probability in Terms of the Electromagnetic Green Tensor: First-Principles Derivation of Equation\ \ref{EELSQM}} We start from eq\ \ref{P0n} for the probability $\Gamma_n^0$ of exciting a mode $n$, which is in turn derived below. The spectrally resolved EELS probability is then given by
\begin{align}
\Gamma_{\rm EELS}(\omega)&=\sum_n\,\Gamma_n^0\,\delta(\omega-\omega_{n0}) \nonumber\\
&=\int d^3 \rb\,|\psi^0(\rb)|^2 \sum_n |\tilde{\beta}_n(\Rb)|^2 \delta(\omega -\omega_{n0}),
\label{PEELSintermediate}
\end{align}
where $\tilde{\beta}_n(\Rb)$ is defined in eq\ \ref{betan}. Starting from the Dirac equation, we derive an effective Schr\"odinger equation to describe the electron and its interaction with an external light field in the linearized-minimal-coupling and nonrecoil approximations (see details in ref\ \citenum{paper339}).  The interaction Hamiltonian then reduces to
\begin{align}
\hat{\mathcal{H}}_1(\rb)=\frac{e\vb}{c}\cdot\hat{\Ab}(\rb),
\label{H1}
\end{align}
where $\hat\Ab(\rb)$ is the vector potential operator, using a gauge in which the scalar potential vanishes. Inserting eq\ \ref{H1} into eq\ \ref{betan}, and this in turn into eq\ \ref{PEELSintermediate}, we find
\begin{align}
&\Gamma_{\rm EELS}(\omega)=\frac{e^2}{\hbar^2c^2}\int d^3\rb|\psi^0(\rb)|^2 \label{EELSprefinal}\\
&\times\int_{-\infty}^\infty dz' \int_{-\infty}^\infty dz'' \ee^{\ii\omega(z''-z')/v} \nonumber\\
&\times\sum_n \big\langle 0\big|\hat{A}_z(\Rb,z') \big|n\big\rangle\big\langle n\big|\hat{A}_z(\Rb,z'')\big|0\big\rangle\;\delta(\omega-\omega_{n0}),
\nonumber
\end{align}
where we have used the hermiticity of $\hat{\Ab}(\rb)$ and taken $\vb=v\zz$. This result can be expressed in terms of the electromagnetic Green tensor, implicitly defined in eq\ \ref{Green} for local media (and by an analogous relation when including nonlocal effects \cite{paper357}), by using the identity (see below)
\begin{align}
&\sum_n \big\langle 0\big|\hat{A}_z(\rb)|n\rangle\langle n|\hat{A}_z(\rb')\big|0\big\rangle \delta(\omega-\omega_{n0}) \nonumber\\
&=-4\hbar c^2\, {\rm Im}\{G_{zz}(\rb,\rb',\omega)\},
\label{AAG}
\end{align}
which is valid for reciprocal materials held at zero temperature, with $n=0$ referring to the sample ground state. Combining eqs\ \ref{EELSprefinal} and \ref{AAG}, we find
\begin{align}
&\Gamma_{\rm EELS}(\omega)=\frac{4e^2}{\hbar}\int d^3\rb|\psi^0(\rb)|^2 \nonumber\\
&\times\int_{-\infty}^\infty dz' \int_{-\infty}^\infty dz'' \cos\left[\omega(z''-z')/v\right]\nonumber\\
&\times{\rm Im}\{-G_{zz}(\Rb,z',\Rb,z'',\omega)\},
\nonumber
\end{align}
where we have transformed $\ee^{\ii\omega(z''-z')/v}$ into a cosine function by exploiting the reciprocity relation $G_{zz}(\rb,\rb',\omega)=G_{zz}(\rb',\rb,\omega)$. Finally, eq\ \ref{EELSQM} is obtained by considering an electron wave function that is tightly confined around a lateral position $\Rb=\Rb_0$ ({\it i.e.}, for $\int_{-\infty}^\infty dz\,|\psi^0(\rb)|^2\approx\delta(\Rb-\Rb_0))$.

\subsection{Derivation of Equation\ \ref{AAG}} Starting with the definition of the retarded electromagnetic Green tensor in a gauge with zero scalar potential at zero temperature,
\begin{align}
G^{\rm R}_{aa'}({\bf r},{\bf r}',t-t')=-\frac{\ii}{4\pi\hbar c^2}\langle0|[\hat{A}_a({\bf r},t),\hat{A}_{a'}({\bf r'},t')]|0\rangle \theta(t-t'),
\nonumber
\end{align}
where $a$ and $a'$ denote Cartesian components, whereas $\theta$ is the step function, we introduce a complete set of eigenstates $|n\rangle$ of the light+matter Hamiltonian $\hat{\mathcal{H}}_0$ ({\it i.e.}, $\hat{\mathcal{H}}_0|n\rangle=\hbar\omega_n|n\rangle$), use the relation $\hat{\Ab}({\bf r},t)=\ee^{\ii\hat{\mathcal{H}}_0t/\hbar}\hat{\Ab}({\bf r})\ee^{-\ii\hat{\mathcal{H}}_0t/\hbar}$ between operators in the Schr\"odinger and Heisenberg pictures, and apply the integral $\int_0^\infty dt \,\ee^{\ii s t}=\ii/(s+\ii 0^+)$ to write \cite{BT1962}
\begin{align}
&G^{\rm R}_{aa'}({\bf r},{\bf r}',\omega) \label{greenw} \\
&=\frac{1}{4\pi \hbar c^2}\int_0^\infty d\omega ' \left[\frac{J_{aa'}(\bf r,r',\omega')}{\omega-\omega'+\ii0^+}-\frac{J_{aa'}^*(\bf r,r',\omega')}{\omega+\omega'+\ii0^+}\right],\nonumber
\end{align}
where
\[J_{aa'}(\rb,\rb',\omega)=\sum_n \langle0| \hat{A}_a({\bf r})|n\rangle \langle n|\hat{A}_{a'}({\bf r}')|0\rangle \delta(\omega -\omega_{n0})\]
is the spectral tensor, $\omega_{n0}=\omega_n-\omega_0$, and $G^{\rm R}({\bf r},{\bf r}',\omega)=\int_{-\infty}^\infty dt\,\ee^{\ii \omega t}\,G^{\rm R}({\bf r},{\bf r}',t)$. The electromagnetic Green tensor in eq\ \ref{greenw} can be shown \cite{AGD1965} to satisfy eq\ \ref{Green} ({\it i.e.}, we have $G^{\rm R}\equiv G$), provided the optical response of the system is assumed to be described by a local, frequency-dependent permittivity $\epsilon(\rb,\omega)$. Now, we introduce the quantum mechanical version of the time-reversal operator $\hat{\Theta}$. Under the assumption of time-reversal symmetry, we have $[\hat{\mathcal{H}}_0,\hat{\Theta}]=0$, and consequently, $\hat{\mathcal{H}}_0|\hat{\Theta} n\rangle=\hbar \omega_n|\hat{\Theta} n\rangle$. Furthermore, assuming a non-degenerate ground state $|0\rangle$, it must obviously satisfy $|\hat{\Theta}0\rangle =|0\rangle$, and therefore, because the time-reversed eigenstates form a complete basis set with the same energies, we can rewrite the spectral tensor as
\begin{align}
&J_{aa'}(\rb,\rb',\omega) \nonumber\\
&=\sum_n\langle\hat{\Theta}0|\hat{A}_a({\bf r})|\hat{\Theta}n\rangle\langle \hat{\Theta}n|\hat{A}_{a'}({\bf r}')|\hat{\Theta}0\rangle\delta(\omega-\omega_{n0}).
\nonumber
\end{align}
Then, using the relation \cite{S1994} $\langle n|\hat{O}|n'\rangle^*=\pm\langle \hat{\Theta} n|\hat{O}|\hat{\Theta} n'\rangle$, which is valid for any Hermitian operator $\hat{O}$ ({\it e.g.}, with $-$ for $\hat{O}=\hat{\Ab}$), we find that $J(\rb,\rb',\omega)=J^*(\rb,\rb',\omega)$ is real. Finally, taking the imaginary part of eq\ \ref{greenw} and using the above property of $J$, together with $1/(s+\ii 0^+)=P[1/s]-\ii \pi \delta(s)$, we obtain $J_{aa'}(\rb,\rb',\omega)=-4\hbar c^2\,{\rm Im}\left\{G_{aa'}({\bf r},{\bf r}',\omega)\right\}$, which reduces to eq\ \ref{AAG} for $a=a'=z$.

\subsection{Inelastic Electron Scattering at Finite Temperature: Derivation of Equation\ \ref{EELST}} The large kinetic energy of beam electrons allows us to safely distinguish them from other electrons in the sample. A free electron initially prepared in state $\qb$ can experience transitions to final states $\qb'$ accompanied by excitations $i$ in the sample. The most general Hamiltonian that describes this interaction, assuming linear coupling to the sample and neglecting electron spin-flips, can be written as
\begin{align}
\hat{\mathcal{H}}_1=\sum_{i\qb\qb'}c^\dagger_{\qb'}c_\qb\,\left(V_{i\qb\qb'}a_i+V_{i\qb'\qb}^*a^\dagger_i\right),
\nonumber
\end{align}
where $a_i$ and $c_\qb$ ($a^\dagger_i$ and $c^\dagger_\qb$) annihilate (create) an excitation $i$ and an electron in state $\qb$, respectively. The label $i$ runs over all possible modes in the system, including plasmons, excitons, phonons, and photons in the radiation field. The details of the interaction are fully contained in the coupling coefficients $V_{i\qb\qb'}$. Within the linear response approximation, and assuming the sample to be initially prepared in thermal equilibrium at temperature $T$, we can write the transition rate between $\qb$ and $\qb'$ electron states using the Fermi golden rule as
\begin{align}
&P_{\qb'\qb}=\frac{2\pi}{\hbar^2}\frac{1}{Z}\sum_{\{n'_i\}}\sum_{\{n_i\}}\exp\left(-\sum_i n_i\,\frac{\omega_i}{\omega_T}\right) \nonumber\\
&\times\left|\left\langle \qb',\{n'_i\}|\hat{\mathcal{H}}_1|\qb,\{n_i\}\right\rangle\right|^2\;\delta\big[\varepsilon_{\qb'}-\varepsilon_\qb+\sum_i(n'_i-n_i)\omega_i\big],
\nonumber
\end{align}
where $\omega_T=\kB T/\hbar$ is the thermal frequency, $\{n_i\}$ describes the initial state of the system through the occupation numbers $n_i$ of modes $i$ having energies $\hbar\omega_i$; the sum in $\{n'_i\}$ runs over all possible final occupations; we introduce the partition function $Z\equiv\sum_{\{n_i\}}\exp(-\sum_in_i\omega_i/\omega_T)=\prod_i\sum_{n_i}\ee^{-n_i\omega_i/\omega_T}$, which allows us to weight each initial configuration $\{n_i\}$ by $Z^{-1}\exp(-\sum_in_i\omega_i/\omega_T)$ (its statistical probability at temperature $T$); and the electron initial and final energies are denoted $\hbar\varepsilon_\qb$ and $\hbar\varepsilon_{\qb'}$, respectively. Now, given the linear dependence of $\hat{\mathcal{H}}_1$ on the operators $a_i$ and $a^\dagger_i$, the initial and final occupation numbers within each term of the sum in $P_{\qb'\qb}$ must differ only for a single $i$, with $n'_i=n_i\pm1$. We can factor out all other $i$'s and separate the rate in energy loss ($n'_i=n_i+1$) and gain ($n'_i=n_i-1$) contributions to write $P_{\qb'\qb}=\int_0^\infty d\omega\,P_{\qb'\qb}(\omega)$, where
\begin{align}
P_{\qb'\qb}(\omega)=N^+(\omega)P_{\qb'\qb,0}^+(\omega)+N^-(\omega)P_{\qb'\qb,0}^-(\omega)
\label{PP}
\end{align}
is the spectrally resolved transition rate,
\begin{align}
P_{\qb'\qb,0}^\pm(\omega)=\frac{2\pi}{\hbar^2}\sum_i|V_{i\qb'\qb}|^2\;\delta(\omega-\omega_i)\;\delta\left[(\qb'-\qb)\cdot\vb\pm\omega\right]
\label{PP0}
\end{align}
are temperature-independent loss (+) and gain (-) rates, and
\[N^\pm(\omega)=\frac{\sum_{n_i}\ee^{-n_i\omega/\omega_T}\,\left|\left\langle n_i\pm1|a^\dagger_i+a_i|n_i\right\rangle\right|^2}{\sum_{n_i}\ee^{-n_i\omega/\omega_T}}.\]
In the derivation of these expressions, we have adopted the nonrecoil approximation for the electron energy difference (eq\ \ref{nonrecoil}) and assumed the condition $\sum_{i}|V_{i\qb\qb'}|^2\delta(\omega-\omega_i)=\sum_{i}|V_{i\qb'\qb}|^2\delta(\omega-\omega_i)$ for each partial sum restricted to degenerate modes $i$ sharing a common frequency $\omega$. This condition, which is satisfied in reciprocal media, also renders $\sum_{\qb'}P_{\qb'\qb,0}^-(\omega)=\sum_{\qb'}P_{\qb'\qb,0}^+(\omega)$ after summing over final states $\qb'$. Finally, we obtain the EELS probability $\Gamma_{\rm EELS}$ by dividing the rates $P$ by the electron current.

For bosonic excitations ({\it e.g.}, photons, phonons, and plasmons), we have $|\langle n_i+1|a^\dagger_i+a_i|n_i\rangle|^2=n_i+1$ and $|\langle n_i-1|a^\dagger_i+a_i|n_i\rangle|^2=n_i$, which allow us to carry out the $n_i$ sums to find $N^+(\omega)=n_T(\omega)+1$ and $N^-(\omega)=n_T(\omega)$, where
\begin{align}
n_T(\omega)=\frac{1}{\ee^{\omega/\omega_T}-1}
\label{BE}
\end{align}
is the Bose-Einstein distribution function. Using these elements in combination with eqs\ \ref{PP} and \ref{PP0}, we directly obtain eq\ \ref{EELST} for the relation between the finite- and zero-temperature EELS probabilities.

For Fermionic excitations ({\it e.g.}, two-level atoms), $n_i$ can take the values 0 or 1, so we have instead $N^+(\omega)=1-n_T^{\rm F}(\omega)$ and $N^-(\omega)=n_T^{\rm F}(\omega)$, where $n_T^{\rm F}(\omega)=1\big{/}(\ee^{\omega/\omega_T}+1)$ is the Fermi-Dirac distribution function for zero chemical potential. The loss and gain probabilities are then given by eq\ \ref{EELST}, but with $n_T(\omega)$ substituted by $-n_T^{\rm F}(\omega)$.

\subsection{Derivation of Equation\ \ref{Gammafi}} For a free electron prepared in an initial monochromatic state $\psi_i(\rb)\ee^{-\ii\varepsilon_it}$ of energy $\hbar\varepsilon_i$, the inelastic scattering probability can be decomposed in contributions arising from transitions to specific final states $\psi_f(\rb)\ee^{-\ii\varepsilon_ft}$. Working within first-order perturbation theory, we consider the transition matrix element
\[\langle fn|\hat{\mathcal{H}}_1|i0\rangle=\frac{ev}{c}\int d^3\rb\;\psi_f^*(\rb)\psi_i(\rb)\,\left\langle n\left|\hat{A}_z(\rb)\right|0\right\rangle\]
for electron-sample transitions driven by the interaction Hamiltonian in eq\ \ref{H1}. From here, Fermi's golden rule yields the transition probability $\Gamma_{fi}=\int_0^\infty d\omega\,\Gamma_{fi}(\omega)$ with
\begin{align}
\Gamma_{fi}(\omega)=\frac{2\pi e^2vL}{\hbar^2c^2}&\sum_n\left|\int d^3\rb\;\psi_f^*(\rb)\psi_i(\rb)\,\left\langle n\left|\hat{A}_z(\rb)\right|0\right\rangle\right|^2 \nonumber\\
&\times\delta(\omega-\omega_{n0})\,\delta(\varepsilon_f-\varepsilon_i+\omega),
\label{gfi}
\end{align}
where $L$ is the quantization length along the e-beam direction and we have multiplied by the interaction time $L/v$ to transform the rate into a probability. Incidentally, this quantity is related to the EELS probability through $\Gamma_{\rm EELS}(\omega)=\sum_f\Gamma_{fi}(\omega)$. Now, expanding the squared modulus in eq\ \ref{gfi} and using eq\ \ref{AAG}, we find
\begin{align}
\Gamma_{fi}(\omega)=\frac{8\pi e^2vL}{\hbar}&\int d^3\rb\int d^3\rb'\;\psi_f(\rb)\psi_i^*(\rb)\psi_f^*(\rb')\psi_i(\rb') \nonumber\\
&\times{\rm Im}\{-G_{zz}(\rb,\rb',\omega)\}
\,\delta(\varepsilon_f-\varepsilon_i+\omega).
\label{Gammafi3D}
\end{align}
Finally, eq\ \ref{Gammafi} is derived from eq\ \ref{Gammafi3D} by factorizing the incident and final electron wave functions as $\psi_{i|f}(\rb)\propto\psi_{i|f\perp}(\Rb)\,\ee^{\ii q_{i|f,z}z}/\sqrt{L}$, summing over final longitudinal wave vectors by means of the prescription $\sum_{q_{f,z}}\rightarrow(L/2\pi)\int_{-\infty}^\infty dq_{f,z}$, and using the $\delta$ function in combination with the nonrecoil approximation $\varepsilon_f-\varepsilon_i\approx(q_{f,z}-q_{i,z})\,v$ (eq\ \ref{nonrecoil}).

\subsection{Derivation of Equation\ \ref{P0n}} We calculate the excitation probability of a sample mode $n$ by tracing out over all final electron states as
\begin{align}
&\Gamma_n^0=\sum_\qb|\alpha_{\qb n}(\infty)|^2 \nonumber\\
&=\frac{(2\pi)^2}{\hbar^2}\sum_\qb\left|\sum_{\qb'}\delta(\varepsilon_\qb-\varepsilon_{\qb'}+\omega_{n0})\,\langle\qb n|\hat{\mathcal{H}}_1|\qb'0\rangle\,\alpha^0_{\qb'}\right|^2,
\label{S1}
\end{align}
where the rightmost expression is obtained by using eq\ \ref{alphainfinity}. We now apply the prescription $\sum_\qb\rightarrow V\int d^3\qb/(2\pi)^3$ to convert electron wave vector sums into integrals, adopt the nonrecoil approximation (eq\ \ref{nonrecoil}), and express the electron part of the matrix element in eq\ \ref{S1} as a real-space integral, using the representation $\langle\rb|\qb\rangle=V^{-1/2}\,\ee^{\ii\qb\cdot\rb}$ for the electron momentum states. Then, taking the electron velocity vector $\vb$ along $\zz$, we obtain $\langle\qb n|\mathcal{H}_1|\qb'0\rangle=V^{-1}\int d^3\rb\;\ee^{\ii(\qb'-\qb)\cdot\rb}\,\langle n|\mathcal{H}_1(\rb)|0\rangle$, and from here
\begin{align}
\Gamma_n^0=\frac{V}{(2\pi)^7\hbar^2v^2}&\int d^3\qb\;\bigg|\int d^3\qb'\,\delta(q_z-q'_z+\omega_{n0}/v) \nonumber\\
&\times\int d^3\rb\;\ee^{\ii(\qb'-\qb)\cdot\rb}\,\langle n|\hat{\mathcal{H}}_1(\rb)|0\rangle\,\alpha^0_{\qb'}\bigg|^2.
\nonumber
\end{align}
We can use the $\delta$ function to perform the $q'_z$ integral and then change the integration variable from $q_z$ to $q_z+\omega_{n0}/v$, so $\Gamma_n^0$ becomes
\begin{align}
\Gamma_n^0=\frac{V}{(2\pi)^7\hbar^2v^2}&\int d^3\qb\bigg|\int d^2\Rb\;\ee^{-\ii\qb_\perp\cdot\Rb}\int d^2\qb'_\perp \,\alpha^0_{(\qb'_\perp,q_z)} \nonumber\\
&\times\ee^{\ii\qb'_\perp\cdot\Rb}\int_{-\infty}^\infty dz\;\ee^{\ii\omega_{n0}z/v}\,\langle n|\hat{\mathcal{H}}_1(\rb)|0\rangle\ \bigg|^2,
\nonumber
\end{align}
where we adopt the notation $\rb=(\Rb,z)$ and $\qb=(\qb_\perp,q_z)$ with $\Rb$ and $\qb_\perp$ standing for real-space and wave-vector coordinate components in the plane perpendicular to the beam direction. This expression can be simplified using the relation $\int d^2\qb_\perp \left|\int d^2\Rb\,\ee^{-\ii\qb_\perp\cdot\Rb}f(\Rb)\right|^2=(2\pi)^2\int d^2\Rb\,|f(\Rb)|^2$ and then changing $\qb'_\perp$ to $\qb_\perp$ to obtain
\begin{align}
\Gamma_n^0=\frac{V}{(2\pi)^5\hbar^2v^2}&\int d^2\Rb\left[\int_{-\infty}^\infty dq_z\left|\int d^2\qb_\perp \,\alpha^0_\qb\,\ee^{\ii\qb_\perp\cdot\Rb}\right|^2\right]\nonumber\\
&\times\left[\left|\int_{-\infty}^\infty dz\;\ee^{\ii\omega_{n0}z/v}\,\langle n|\hat{\mathcal{H}}_1(\rb)|0\rangle\ \right|^2\right].
\nonumber
\end{align}
Finally, using the identity $\int_{-\infty}^\infty dq_z\left|\int d^2\qb_\perp \,\alpha^0_\qb\,\ee^{\ii\qb_\perp\cdot\Rb}\right|^2
=(2\pi)^{-1}\int_{-\infty}^\infty dz\left|\int d^3\qb \,\alpha^0_\qb\,\ee^{\ii\qb\cdot\rb}\right|^2$, we find the result 
\begin{align}
\Gamma_n^0=&\int d^3\rb\left[\left|V^{1/2}\int\frac{d^3\qb}{(2\pi)^3}\,\alpha^0_\qb\,\ee^{\ii\qb\cdot\rb}\right|^2\right]\nonumber\\
&\times\left[\left|\frac{1}{\hbar v}\int_{-\infty}^\infty dz\;\ee^{-\ii\omega_{n0}z/v}\,\langle 0|\hat{\mathcal{H}}_1(\rb)|n\rangle\ \right|^2\right],
\label{Pn0final}
\end{align}
which reduces to eq\ \ref{P0n} with $\psi^0(\rb)$ and $\tilde{\beta}_n(\Rb)$ defined by eqs\ \ref{psi0} and \ref{betan}.

\subsection{Derivation of Equations\ \ref{PNn}-\ref{Mnj}} A direct extension of the general formalism used in the previous paragraph allows us to deal with $N$ free  independent electrons prepared in initial states (before interaction with the sample) described by their wave function coefficients $\alpha_\qb^j$ with $j=0,\dots,N-1$. The wave function of the combined system formed by the sample and the electrons can be written as
\begin{align}
|\psi(t)\rangle=\sum_{\{\qb\} n}\alpha_{\{\qb\} n}(t)\ee^{-\ii\left(\sum_j\varepsilon_{\qb_j}+\omega_{n0}\right)t}|\{\qb\}n\rangle,
\nonumber
\end{align}
where $\{\qb\}$ denotes the ensemble of wave vectors $\qb_j$. Given the large size of the electron configuration space in a microscope, we consider that it is safe to disregard spin degrees of freedom and the Pauli exclusion principle ({\it i.e.}, we consider distinguishable electrons). We further neglect electron-electron Coulomb interaction in the beam. Additionally, we work in the weak coupling regime, under the assumption that the sample is excited once at most by the passage of the $N$ electrons, which is a good approximation for $N\ll 1/\Gamma_n^0$ (we note that typical excitation probabilities are $\Gamma_n^0\lesssim10^{-5}$ per electron for single sample modes $n$). This allows us to integrate the Schr\"odinger equation to find the wave function coefficients after interaction as a generalization of eq\ \ref{alphainfinity}:
\begin{align}
\alpha_{\{\qb\}n}(\infty)=-\frac{2\pi\ii}{\hbar}&\sum_{\{\qb'\}}\delta\left(\omega_{n0}+{\sum}_j\varepsilon_{\qb_j\qb'_j}\right)\nonumber\\
&\times\langle\{\qb\}n|\hat{\mathcal{H}}_1|\{\qb'\}0\rangle\prod_j\alpha^j_{\qb'_j},
\label{alphanN}
\end{align}
where $\varepsilon_{\qb_j\qb'_j}=\varepsilon_{\qb_j}-\varepsilon_{\qb'_j}$. Now, each of the terms in the real-space representation of the interaction Hamiltonian $\hat{\mathcal{H}}_1({\rb})=\sum_j\hat{\mathcal{H}}_1(\rb_j)$ depends on just one of the electron coordinates, and thus, because of the orthogonality of the electron momentum states, $\{\qb\}$ and $\{\qb'\}$ in eq\ \ref{alphanN} differ by no more than one of the electron wave vectors. This allows us to recast eq\ \ref{alphanN} as
\begin{align}
\alpha_{\{\qb\}n}(\infty)=-\frac{2\pi\ii}{\hbar }&\left({\prod}_j\alpha^j_{\qb_j}\right)\sum_{j}\sum_{\qb'_j}\delta\left(\omega_{n0}+\varepsilon_{\qb_j\qb'_j}\right)\nonumber\\
&\times\langle\qb_jn|\hat{\mathcal{H}}_1|\qb'_j0\rangle\left(\alpha^j_{\qb'_j}/\alpha^j_{\qb_j}\right).
\label{alphanNbis}
\end{align}
The excitation probability of sample mode $n$ is obtained by tracing out the final electron states as
\begin{align}
\Gamma_n^{\rm total}=\sum_{\{\qb\}}\left|\alpha_{\{\qb\}n}(\infty)\right|^2,
\label{PnN0}
\end{align}
which, in combination with eq\ \ref{alphanNbis} and the normalization condition of the initial states $\sum_\qb\left|\alpha^j_\qb\right|^2=1$, leads to (eq\ \ref{PNn})
\begin{align}
\Gamma_n^{\rm total}=\sum_j \Gamma_n^j + \sum_{j\neq j'} Q_n^jQ_n^{j'*},
\label{Pntotal}
\end{align}
where
\begin{align}
\Gamma_n^j=\frac{(2\pi)^2}{\hbar^2}\sum_{\qb_j}\left|\sum_{\qb'_j}\delta\left(\varepsilon_{\qb_j\qb'_j}+\omega_{n0}\right)\langle\qb_jn|\hat{\mathcal{H}}_1|\qb'_j0\rangle\,\alpha^j_{\qb'_j}\right|^2
\label{Pnj}
\end{align}
and
\begin{align}
Q_n^j=\frac{2\pi}{\hbar}\sum_{\qb_j\qb'_j}\delta\left(\varepsilon_{\qb_j\qb'_j}+\omega_{n0}\right)\langle\qb'_j0|\hat{\mathcal{H}}_1|\qb_jn\rangle\,\alpha^{j*}_{\qb'_j}\alpha^j_{\qb_j}.
\label{Pnjj}
\end{align}
Noticing that eq\ \ref{Pnj} is just like eq\ \ref{S1} with $\alpha^0_{\qb'}$ substituted by $\alpha^{j}_{\qb'_j}$, we can write from eq\ \ref{Pn0final}
\begin{align}
\Gamma_n^j=\int d^3\rb\;|\psi^j(\rb)|^2 |\tilde{\beta}_n(\Rb)|^2
\label{Pnjfinal}
\end{align}
with
\begin{align}
\psi^j(\rb)=V^{1/2}\int \frac{d^3\qb}{(2\pi)^3}\,\alpha^j_\qb\,\ee^{\ii\qb\cdot\rb}.
\label{psij}
\end{align}
Now, using the nonrecoil approximation $\varepsilon_{\qb_j\qb'_j}=v(q_{jz}-q'_{jz})$, transforming wave vector sums into integrals, expressing matrix elements as real-space integrals, and proceeding in a similar way as in the derivation of eq\ \ref{Pn0final}, we can rearrange eq\ \ref{Pnjj} as
\begin{align}
Q_n^j=\frac{V}{(2\pi)^5\hbar v}&\int d^3\qb_j\int d^3\qb'_j\int d^2\Rb\;\ee^{\ii(\qb_{j\perp}-\qb'_{j\perp})\cdot\Rb}\nonumber\\
&\times\alpha^{j*}_{\qb'_j}\alpha^j_{\qb_j} \; \delta(q_{jz}-q'_{jz}+\omega_{n0}/v) \nonumber\\
&\times\int_{-\infty}^\infty dz\;\ee^{-\ii\omega_{n0}z/v}\langle0|\hat{\mathcal{H}}_1(\rb)|n\rangle \label{Qnj}
\end{align}
which reduces to eq\ \ref{Qj} with $\tilde{\beta}_n(\Rb)$ defined in eq\ \ref{betan}, whereas
\begin{align}
M_n^j(\Rb)&=\frac{V}{(2\pi)^{5}}\int d^3\qb_j\int d^3\qb'_j\;\ee^{\ii(\qb_{j\perp}-\qb'_{j\perp})\cdot\Rb}\nonumber\\
&\quad\quad\quad\times\alpha^j_{\qb_j}\alpha^{j*}_{\qb'_j}\;\delta(q_{jz}-q'_{jz}+\omega_{n0}/v) \nonumber\\
&=\frac{V}{(2\pi)^{5}}\int d^3\qb_j\int d^3\qb'_j\;\ee^{\ii(\qb_{j\perp}-\qb'_{j\perp})\cdot\Rb}\nonumber\\
&\quad\quad\quad\times\alpha^j_{\qb_j}\alpha^{j*}_{\qb'_j}\;\frac{1}{2\pi}\int_{-\infty}^\infty dz \;\ee^{\ii(q_{jz}-q'_{jz}+\omega_{n0}/v)z} \nonumber\\
&=\int_{-\infty}^\infty dz \;\ee^{\ii\omega_{n0}z/v} \bigg[V^{1/2}\int \frac{d^3\qb_j}{(2\pi)^3}\,\alpha^j_{\qb_j}\;\ee^{\ii\qb_j\cdot\rb}\bigg]\nonumber\\
&\quad\quad\quad\quad\quad\times\bigg[V^{1/2}\int \frac{d^3\qb'_j}{(2\pi)^3}\,\alpha^{j*}_{\qb'_j}\;\ee^{-\ii\qb'_j\cdot\rb}\bigg] \nonumber\\
&=\int_{-\infty}^\infty dz \;\ee^{\ii\omega_{n0}z/v}\;|\psi^j(\rb)|^2
\nonumber
\end{align}
becomes eq\ \ref{Mnj}, the Fourier transform of the electron probability density in the incident electron wave function $j$.

\subsection{Derivation of Equations\ \ref{Pnlocal1} and \ref{Pnlocal2}} We consider electron wave functions constructed in terms of normalized Gaussian wavepackets of the form $\psi_G(\rb)=\psi_\perp(\Rb)\,\ee^{-z^2/2\Delta^2}/\pi^{1/4}\Delta^{1/2}$, where we factorize the transverse dependence in $\psi_\perp(\Rb)$. For simplicity, we approximate $|\psi_\perp(\Rb)|^2\approx\delta(\Rb)$ under the assumption that the transverse width $w$ is small compared with the characteristic length of variation of the electric field associated with the excited mode $n$, or equivalently, $|\nabla_\Rb\tilde{\beta}_n(\Rb)|\ll1/w$. The configurations discussed in Figures\ \ref{Fig5} and \ref{Fig6} involve electron wave functions of the general form
\begin{align}
\psi^j(\rb)=N_j^{-1}\sum_s\gamma_s^j\psi_G(\rb-\rb_s),
\label{psi11}
\end{align}
where we assume the same longitudinal wavepacket width $\Delta$ for all components, and $N_j=\big(\sum_{ss'}\gamma_s^j\gamma_{s'}^{j*}I_{ss'}\big)^{1/2}$ is a normalization constant that depends on the overlap integrals
\begin{align}
I_{ss'}=\left\{\begin{matrix} 
\ee^{-(z_s-z_{s'})^2/4\Delta^2}, & \quad\quad\quad\text{if $\Rb_s=\Rb_{s'}$,} \\ \!\!\!0, & \quad\quad\quad\text{otherwise.}\end{matrix}\right.
\nonumber
\end{align}
Plugging eq\ \ref{psi11} into eqs\ \ref{Pnjfinal} and \ref{Qj}, we readily find
\begin{subequations}
\label{PandQ}
\begin{align}
\Gamma_n^j&=\frac{\sum_{ss'}\gamma_s^j\gamma_{s'}^{j*}I_{ss'}\,\left|\tilde{\beta}_n(\Rb_s)\right|^2}{\sum_{ss'}\gamma_s^j\gamma_{s'}^{j*}I_{ss'}} \nonumber\\
&\approx \frac{\sum_{s}|\gamma_s^j|^2\,\left|\tilde{\beta}_n(\Rb_s)\right|^2}{\sum_{s}|\gamma_s^j|^2},
\label{Pnjsup}\\
Q_n^j&=\sqrt{S}\;\frac{\sum_{ss'}\gamma_s^j\gamma_{s'}^{j*}I_{ss'}\,\ee^{\ii\omega_{n0}(z_s+z_{s'})/2v}\,\tilde{\beta}_n(\Rb_s)}{\sum_{ss'}\gamma_s^j\gamma_{s'}^{j*}I_{ss'}} \nonumber\\
&\approx\sqrt{S}\;\frac{\sum_{s}|\gamma_s^j|^2\,\ee^{\ii\omega_{n0}z_s/v}\,\tilde{\beta}_n(\Rb_s)}{\sum_{s}|\gamma_s^j|^2},
\label{Qnjsup}
\end{align}
\end{subequations}
where
\begin{align}
S 
=\ee^{-\omega_{n0}^2\Delta^2/2v^2}.
\nonumber
\end{align}
The rightmost approximations in eqs\ \ref{PandQ} correspond to the nonoverlapping wavepacket limit ({\it i.e.},  $|z_s-z_{s'}|\gg\Delta$ for $s\neq s'$ and $\Rb_s=\Rb_{s'}$), which yields $I_{ss'}=\delta_{s,s'}$. Now, we adopt this limit and specify eqs\ \ref{Pntotal} and \ref{PandQ} for the beams studied in Figures\ \ref{Fig5} and \ref{Fig6}:
\begin{itemize}
\item {\bf Figure\ \ref{Fig5}b.}
(1) We consider two Gaussian wavepackets $s=0,1$ with longitudinal coordinates $z_0=0$ and $z_1=a$, where $a\gg\Delta$ is the wavepacket separation, and the same lateral coordinates $\Rb_s=\bb$, so $\tilde{\beta}_n(\Rb_s)=\tilde{\beta}_n(\bb)$ is independent of $s$ and factors out in eqs\ \ref{PandQ}; in particular, eq\ \ref{Pnjsup} reduces to $\Gamma_n^j=\left|\tilde{\beta}_n(\bb)\right|^2$.
(2) For two electrons $j=0,1$, each of them fully contained in one of the two wavepackets, we have $|\gamma_s^j|^2=\delta_{s,j}$, so eq\ \ref{Qnjsup} gives $Q_n^0=\sqrt{S}\tilde{\beta}_n(\bb)$ and $Q_n^1=\sqrt{S}\tilde{\beta}_n(\bb)\ee^{\ii\omega_{n0}a/v}$; inserting these expressions in eq\ \ref{Pntotal}, we find $\Gamma_n^{\rm total}=2\left|\tilde{\beta}_n(\bb)\right|^2\left[1+S\,\cos(\omega_{n0}a/v)\right]$ ({\it i.e.}, eq\ \ref{Pnlocal1} with the $+$ sign) (incidentally, this result remains unchanged even when the wavepackets overlap).
(3) If each of the two electrons is equally shared among the two wavepackets, we have $|\gamma_s^j|^2=1/2$; evaluating eq\ \ref{Qnjsup} with these coefficients, we find $Q_n^0=Q_n^1=\sqrt{S}\tilde{\beta}_n(\bb)\left(1+\ee^{\ii\omega_{n0}a/v}\right)/2$, which together with eq\ \ref{Pntotal} lead to the result $\Gamma_n^{\rm total}=2\left|\tilde{\beta}_n(\bb)\right|^2\,\left[1+S\cos^2(\omega_{n0}a/2v)\right]$ ({\it i.e.}, eq\ \ref{Pnlocal2}).
\item {\bf Figure\ \ref{Fig5}c.}
(1) We consider two wavepackets $s=0,1$ with $\Rb_0=-\Rb_1=\bb$, $z_0=0$, and $z_1=a$; because $\left|\tilde{\beta}_n(\Rb_s)\right|$ is also independent of $s$ (see below), we can factor it out in eq\ \ref{Pnjsup}, thus leading again to $\Gamma_n^j=\left|\tilde{\beta}_n(\bb)\right|^2$.
(2) To describe two electrons, each of them separated in different wavepackets, we take $|\gamma_s^j|^2=\delta_{s,j}$, so eq\ \ref{Qnjsup} yields $Q_n^0=\sqrt{S}\tilde{\beta}_n(\bb)$ and $Q_n^1=-\sqrt{S}\tilde{\beta}_n(\bb)\ee^{\ii\omega_{n0}a/v}$, where we have used the property $\tilde{\beta}_n(-\bb)=-\tilde{\beta}_n(\bb)$ for the coefficient of coupling to an excitation with the transition dipole oriented as shown in Figure\ \ref{Fig5}; we thus find from eq\ \ref{Pntotal} the result $\Gamma_n^{\rm total}=2\left|\tilde{\beta}_n(\bb)\right|^2\,\left[1-S\,\cos(\omega_{n0}a/v)\right]$ ({\it i.e.}, eq\ \ref{Pnlocal1} with the $-$ sign).
(3) Proceeding as above for the configuration in which each of the two electrons is equally shared among the two wavepackets, we find $Q_n^0=Q_n^1=\sqrt{S}\tilde{\beta}_n(\bb)\left(1-\ee^{\ii\omega_{n0}a/v}\right)/2$, which now results in $\Gamma_n^{\rm total}=2\left|\tilde{\beta}_n(\bb)\right|^2\,\left[1+S\sin^2(\omega_{n0}a/2v)\right]$ ({\it i.e.}, eq\ \ref{Pnlocal2} with cos replaced by sin).
\item  {\bf Figure\ \ref{Fig6}.} In this configuration, the coupling coefficient has the same spatial periodicity as the excited mode ({\it i.e.}, $\tilde{\beta}_n(\Rb_s)=\tilde{\beta}_n(0)\,\ee^{\ii\kb_{n\parallel}\cdot\Rb_s}$ picks up the mode propagation phase at the region of electron-sample interaction). With the same choice of wave function coefficients as in the above analysis of Figure\ \ref{Fig5}c, and considering a lateral separation $\bb=\Rb_0-\Rb_1$ between the two wavepackets, we straightforwardly find the same expressions for the excitation probability as in Figure\ \ref{Fig5}b, but with $\omega_{n0}a/v$ replaced by $\omega_{n0}a/v-\kb_n\cdot\bb$.
\end{itemize}

In the main text, we also discuss a generalization of Figure\ \ref{Fig5}b to a beam consisting of $N$ electrons ($j=0,\dots,N-1$), each of them distributed among $L$ periodically arranged wavepackets ($s=0,\dots,L-1$) with longitudinal spacing $a$ and the same lateral position $\Rb_s=\bb$ for all. Proceeding in a similar way as in the above analysis of Figure\ \ref{Fig5}b, we take $|\gamma_s^j|^2=1$ and find from eqs\ \ref{PandQ} the results $\Gamma_n^j=\left|\tilde{\beta}_n(\bb)\right|^2$ and $Q_n^j=\sqrt{S}\left|\tilde{\beta}_n(\bb)\right|^2(1/L)\sum_s\ee^{\ii s\omega_{n0}a/v}$, which, combined with eq\ \ref{Pntotal}, leads to eq\ \ref{Gtotn}.



\section*{Acknowledgments} 
We thank Fabrizio Carbone, Archie Howie, Ido Kaminer, Ofer Kfir, Mathieu Kociak, Albert Polman, Claus Ropers, Nahid Talebi, and Jo Verbeeck for helpful and enjoyable discussions. This work has been supported in part by the European Research Council (Advanced Grant 789104-eNANO), the European Commission (Horizon 2020 Grants FET-Proactive 101017720-EBEAM and FET-Open 964591-SMART-electron), the Spanish MINECO (MAT2017-88492-R and Severo Ochoa CEX2019-000910-S), the Catalan CERCA Program, and Fundaci\'{o}s Cellex and Mir-Puig. V.D.G. acknowledges financial support from the EU (Marie Sk\l{}odowska-Curie Grant 713729).



\end{document}